\documentclass[%
 reprint,
superscriptaddress,
 amsmath,amssymb,
 aps,
 prl
]{revtex4-2}

\usepackage{graphicx}
\usepackage{dcolumn}
\usepackage{bm}
\usepackage{hyperref}
\hypersetup{
	colorlinks   = true, 
	urlcolor     = blue, 
	linkcolor    = blue, 
	citecolor    = blue 
}
\usepackage{pifont}
\usepackage{multirow}
\usepackage{tabularray}
\usepackage{array}
    \newcolumntype{P}[1]{>{\centering\arraybackslash}p{#1}}
    \newcolumntype{M}[1]{>{\centering\arraybackslash}m{#1}}
\usepackage{stackengine}
    
\usepackage{romannum}
\usepackage{bigints}
\usepackage{ulem}
\usepackage{xcolor}
\usepackage{tabularx} 
\usepackage{siunitx} 
\usepackage[footnotehyper]{nicematrix}
\NiceMatrixOptions{cell-space-limits = 1.8pt}

\begin{document}

\preprint{APS/123-QED}

\title{Correlated Quantum Dephasometry: Symmetry-Resolved Noise Spectroscopy of Two-Dimensional Superconductors and Altermagnets} 
\author{Wenbo Sun}
\affiliation{Elmore Family School of Electrical and Computer Engineering, Birck Nanotechnology Center, Purdue University, West Lafayette, Indiana 47907, USA}
\author{Zubin Jacob}
\email{zjacob@purdue.edu}
\affiliation{Elmore Family School of Electrical and Computer Engineering, Birck Nanotechnology Center, Purdue University, West Lafayette, Indiana 47907, USA}

\date{\today}

\begin{abstract}
Symmetry-resolved spectroscopies, such as angle-resolved photoemission spectroscopy and polarization-resolved Raman, are central for quantum material characterization, yet remain challenging at nanoscale dimensions and low frequencies. Here, we propose correlated quantum dephasometry, which enables symmetry resolved quantum noise spectroscopy of materials at nanoscale and low ($\sim$MHz) frequencies via correlated dephasing of two spin qubits near materials. Our approach leverages the finite-range spatial structures of nonlocal near-field noise correlations to isolate rotational symmetry of the material response in momentum space beyond single qubit capabilities. We apply our approach to two-dimensional (2D) superconductors, and predict clear fingerprints that discriminate s-, d-, and g-wave symmetry of the superconducting gap. To highlight the generality, we further show that the same framework resolves 2D s-wave antiferromagnets and d-wave altermagnets. Our results establish correlated quantum dephasometry as a nanoscale, low-frequency complement for symmetry-resolved spectroscopy applicable to a broad class of quantum materials.
\end{abstract}

\maketitle

\normalem

\emph{Introduction.--}
Symmetry- and angle-resolved spectroscopies are central to diagnosing quantum materials, from unconventional superconductors to anisotropic magnets. Established techniques such as angle-resolved photoemission spectroscopy (ARPES)~\cite{sobota2021angle,zhang2022angle} and polarization-resolved Raman spectroscopy (PRRS)~\cite{devereaux2007inelastic,pimenta2021polarized} have achieved great success in probing wavevector resolved information about materials, from symmetry-selective excitations to momentum-dependent electronic band structures. However, these methods are typically implemented in mesoscopic settings ($\sim$$\mu m$), and probe comparatively high-frequency phenomena ($>$THz)~\cite{sobota2021angle,devereaux2007inelastic}, which can limit their capabilities to access key signatures of materials at low frequencies (e.g., $\sim$ MHz) or at nanoscale dimensions.

Quantum impurity sensors, e.g., nitrogen-vacancy (NV) centers in diamond and hBN defect centers, provide a natural platform for probing condensed matter systems at low frequencies (e.g., MHz-GHz) with nanoscale spatial resolution~\cite{barry2020sensitivity,degen2017quantum,casola2018probing,gottscholl2021spin,vaidya2023quantum}. Most material noise sensing efforts have focused on the single-qubit level, where single spin relaxation (relaxometry) and dephasing (dephasometry) probe spatially local magnetic field fluctuations (noise) near conductors and magnets~\cite{kolkowitz2015probing,li2024observation,bhattacharyya2024imaging,machado2023quantum,flebus2018quantum,bittencourt2025quantum,dolgirev2022characterizing,andrich2017long,du2017control,pelliccione2016scanned,thiel2016quantitative,mahmud2026probing,ziffer2024quantum,curtis2024probing,liu2025quantum,rovny2024nanoscale,finco2021imaging,chatterjee2019diagnosing,wang2022noninvasive}. This local magnetic noise is primarily connected to wavevector angle averaged material responses, with limited sensitivity to symmetry- or angle-resolved material properties in momentum space.

Recent experiments and theories explored multiqubit sensing of spatially correlated noise~\cite{rovny2022nanoscale,rovny2025multi,curtis2025non,le2025wideband,wu2025spin,gao2025signal,wang2024exponential,prabhu2026exponential}. Considerable efforts focus on global or unstructured correlations in quantum sensor networks and treat global correlations as a metrological resource for enhancing sensitivity~\cite{wang2024exponential,prabhu2026exponential,jeske2014quantum,brady2026correlated}. However, in the near field of materials, fluctuations are neither purely local nor globally correlated, but exhibit finite-range correlations and spatial structures sensitive to material dynamics~\cite{sun2025nanophotonic}. Beyond sensitivity enhancement, leveraging these finite-range spatial structures of near-field correlations to extract material properties beyond single-qubit sensing remains largely unexplored.

In this Letter, we propose correlated quantum dephasometry, which enables symmetry resolved quantum noise spectroscopy of materials via correlated dephasing dynamics accessible to two (or more) spin qubit sensors near materials. We leverage spatial structures of nonlocal near-field noise correlations to isolate rotational symmetry of material response in momentum space beyond the single qubit capabilities. Unlike ARPES and PRRS focusing on high frequencies ($>$THz), our approach accesses the low-frequency ($\sim$MHz) components of the material response. As an application, we apply our approach to two-dimensional (2D) superconductors, and predict clear fingerprints that discriminate s-, d-, and g-wave symmetry of the superconducting gap. To highlight the generality, we further show the same framework distinguishes 
2D s-wave antiferromagnets and d-wave altermagnets. Our results establish correlated quantum dephasometry as a route to symmetry-resolved spectroscopy of low-frequency ($\sim$MHz) material response in momentum space for a broad class of quantum materials. 

\begin{figure*}[!t]
    \centering
    \includegraphics[width = 6.4in]{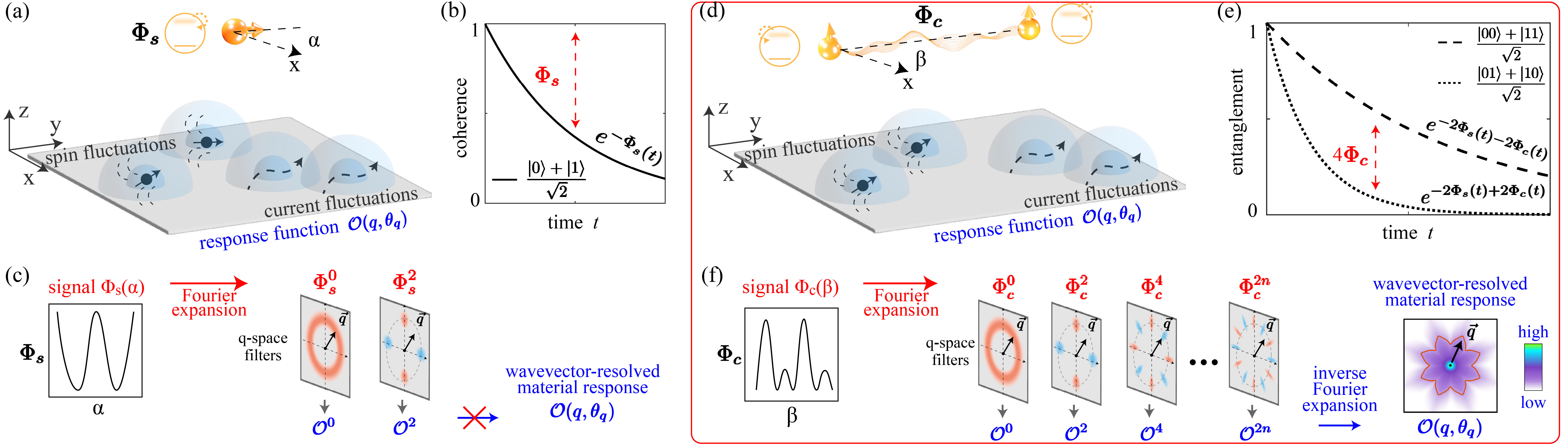}
    \caption{Two-qubit correlated quantum dephasometry enables symmetry-resolved quantum noise spectroscopy by probing spatial noise correlations near quantum materials. (a,b) Single-qubit quantum dephasometry probes local magnetic noise near materials via single qubit dephasing function $\Phi_s(\alpha)$. $\alpha$ is the in-plane orientation of the single qubit. (d,e) Correlated quantum dephasometry probes spatially nonlocal magnetic noise correlations near materials via correlated dephasing function $\Phi_c(\beta)$. $\beta$ represents the two-qubit orientation. (c,f) Correlated quantum dephasometry decomposes the angular symmetry of the nonlocal response function $\mathcal{O}(q,\theta_q)$ in momentum space into rotational Fourier components of $\Phi_c(\beta)$ beyond single-qubit methods.} 
    \label{fig:fig1}
\end{figure*}

\emph{Near-Field Correlated Dephasing.--}We first introduce the near-field correlated dephasing dynamics underlying correlated quantum dephasometry. We consider a pair of spin qubits (e.g., NV centers) at positions $\mathbf{r_1}$, $\mathbf{r_2}$ near 2D materials. Fluctuating currents or magnetizations inside the materials generate fluctuating magnetic fields that couple to the qubits [Fig.~\ref{fig:fig1}(d)] via $\hat{H}_{int}=- \sum_i \mathbf{m}_{i} \hat{\sigma}_i^z \cdot \hat{\mathbf{B}}^{\mathrm{fl}}(\mathbf{r_i}), \, i=1,2$, where $\mathbf{m}_i, \hat{\sigma}_i^z$ are the spin magnetic moment and Pauli operators. Here, spatially local magnetic noise $\langle \hat{\mathbf{B}}^{\mathrm{fl}}(\mathbf{r_i}) \hat{\mathbf{B}}^{\mathrm{fl,\dagger}}(\mathbf{r_i})\rangle$ randomly perturbs qubit frequencies and induces single-qubit dephasing [Fig.~\ref{fig:fig1}(a)]. Meanwhile, in the near-field of materials, magnetic fluctuations at different qubit positions can have prominent spatially nonlocal correlations $\langle \hat{\mathbf{B}}^{\mathrm{fl}}(\mathbf{r_i}) \hat{\mathbf{B}}^{\mathrm{fl,\dagger}}(\mathbf{r_j})\rangle \neq 0$, leading to correlations in dephasing dynamics between spins [Fig.~\ref{fig:fig1}(d)]. The correlated dephasing dynamics are characterized by the correlated dephasing function~\cite{sun2025nanophotonic}: 
\begin{align}
    &\begin{aligned}
          &\Phi_{c}(\mathbf{r_i},\mathbf{r_j},t) =  \int_0^{\omega_c} d\omega \frac{4\mu_0}{\hbar \pi} \coth{\frac{\hbar \omega}{2 k_B T}} F(\omega, t) J_c(\mathbf{r_i},\mathbf{r_j},\omega)
         \label{td_cdf},
    \end{aligned}
\end{align} 
while the single qubit dephasing function is $\Phi_{s}(\mathbf{r_i},t)=\Phi_{c}(\mathbf{r_i},\mathbf{r_i},t)$. In Eq.~(\ref{td_cdf}),  $J_c(\mathbf{r_i},\mathbf{r_j},\omega) \propto \langle \hat{\mathbf{B}}^{\mathrm{fl}}(\mathbf{r_i}) \hat{\mathbf{B}}^{\mathrm{fl,\dagger}}(\mathbf{r_j})\rangle$ is the near-field noise correlation spectrum encoding target material properties. $T$ is the temperature. $F(\omega, t)$ is a frequency filter function controlled by the measurement pulse sequences~\cite{cywinski2008enhance,bar2012suppression}, which sets the noise spectral window contributing to correlated dephasing. 

Unlike single-qubit dephasing $\Phi_s(t)$ typically measured from the coherence decay $e^{-\Phi_s(t)}$ of a qubit initialized in a superposition state [Fig.~\ref{fig:fig1}(b)], correlated dephasing function $\Phi_c(t)$ can be measured from collective two-qubit coherence. For qubits prepared initially in Bell states $|00\rangle + |11\rangle$ ($|01\rangle + |10\rangle$), correlated dephasing interferes constructively (destructively) with single-qubit dephasing, leading to distinct decay of two-qubit entanglement $e^{-\Phi_{|00\rangle + |11\rangle}(t)}$ and $e^{-\Phi_{|01\rangle + |10\rangle}(t)}$~\cite{sun2025nanophotonic},
\begin{align}\label{experimental}
\begin{aligned}
    \Phi_{|00\rangle + |11\rangle}(t)&=\Phi_s(\mathbf{r_i},t)+\Phi_s(\mathbf{r_j},t) + 2\Phi_c(\mathbf{r_i},\mathbf{r_j},t), \\
    \Phi_{|01\rangle + |10\rangle}(t)&=\Phi_s(\mathbf{r_i},t)+\Phi_s(\mathbf{r_j},t)-2\Phi_c(\mathbf{r_i},\mathbf{r_j},t).
\end{aligned}
\end{align}
Therefore, the correlated dephasing function $\Phi_c(t)$ can be isolated from the different disentanglement dynamics of Bell states [Fig.~\ref{fig:fig1}(e)]. Meanwhile, we note that $\Phi_c$ can also be measured without entanglement for spectrally or optically resolved qubits~\cite{le2025wideband,rovny2025multi}. 

\begin{figure*}[!t]
    \centering
    \includegraphics[width=5.7in]{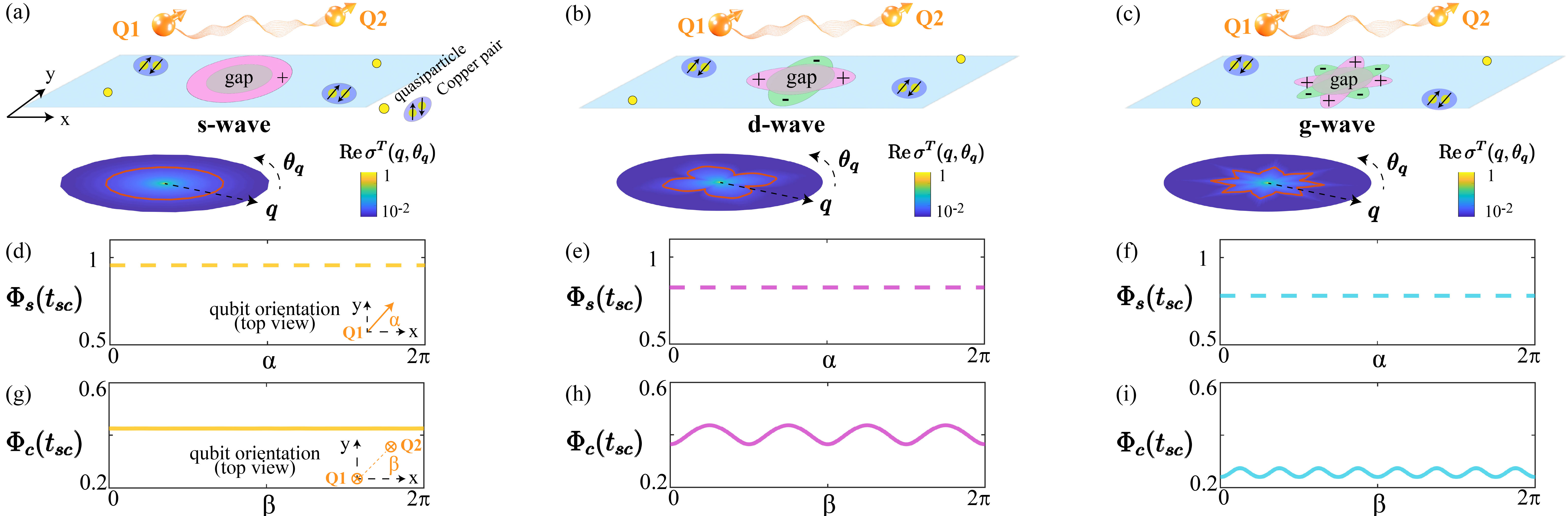}
    \caption{Two-qubit correlated quantum dephasometry isolates the superconductor pairing symmetry. (a-c) Schematic of two spin qubits (Q1, Q2) near 2D superconductors with s-, d-, and g-wave pairing symmetry. The rotational symmetry of transverse conductivity $\mathrm{Re} \, \sigma^T(q,\theta_{q})/\sigma_n$ in momentum space inherits symmetry of gap functions. (d-f) Single-qubit dephasing function $\Phi_s(\alpha)$ exhibits similar behaviors near s-, d-, and g-wave superconductors. (g-i) Correlated dephasing function $\Phi_c(\beta)$ with respect to the two-qubit orientation $\beta$ resolves the pairing symmetry for s-, d, and g-wave superconductors. We take $D=8z$ in (g,h) and $D=12z$ in (i).}
    \label{fig:fig2}
\end{figure*}

\emph{Two-qubit Correlated Dephasometry.--}We now show that the near-field correlated dephasing $\Phi_c$ enables symmetry-resolved quantum noise spectroscopy of material response beyond single qubit methods. For concreteness, we consider two spin qubits with magnetic moment $m_0$ and quantization axes perpendicular to a 2D material. The qubits are positioned at height $z$ above the material and separated by interqubit distance $\mathbf D=\mathbf{r_i}- \mathbf{r_j}$ at angle $\beta$ from the $x$-axis [Fig.~\ref{fig:fig1}(d)]. We prove that general qubit orientations yield the same sensing capabilities in~\cite{SI}. 

The two qubits probe spatial correlations of fluctuating evanescent magnetic fields with in-plane momentum $\mathbf{q}\gg\omega/c$   generated by fluctuating currents/magnetizations inside materials~\cite{sun2025nanophotonic}. With the fluctuation-dissipation theorem (FDT), we can connect near-field magnetic noise correlation spectra $J_c(\mathbf{r_i},\mathbf{r_j},\omega)$ to material response functions $\mathcal{O}$. Here, the spatial nonlocality in $J_c$ introduces an additional phase factor $e^{i\mathbf{q}\cdot \mathbf{D}}$ in the noise correlation spectrum. Expanding this phase factor, we obtain~\cite{SI},
\begin{multline}\label{Jc_sum} J_c(\omega) =\frac{\mu_0 m_0^2}{16\pi^2} \sum_{n=-\infty}^\infty (i)^{2n} e^{i2n\beta} \int_0^\infty dq\, q e^{-2qz} J_{2n}(qD) \\ \int_0^{2\pi}d\theta_q \, e^{-i2n\theta_q} \, \mathcal{O}(q,\theta_q,\omega),
\end{multline}
where $\mathcal{O}$ includes the material response function (e.g., conductivity or magnetic susceptibility, as discussed below), and $J_{2n}$ is the $2n$th Bessel function of the first kind. We note that the angular integral over $\theta_q$ in Eq.~(\ref{Jc_sum}) provides an angular filter function in momentum space that decomposes the material response $\mathcal{O}(q,\theta_q)$ into Fourier coefficients $\mathcal{O}^{2n}$ [Fig.~\ref{fig:fig1}(f)], while $J_{2n}(qD)$ controlled by interqubit distance $D$ sets weights of $\mathcal{O}^{2n}$ in the correlated noise spectrum. 

Substituting Eq.~(\ref{Jc_sum}) into Eq.~(\ref{td_cdf}), we find the correlated dephasing function,
\begin{equation}\label{Phic_Fourier}
\Phi_c(\beta,D,t)=\sum_{n=-\infty}^\infty e^{i2n\beta} \, \Phi^{2n}_c(D,t),
\end{equation}
\begin{align} 
\begin{aligned}
&\Phi^{2n}_c(D)= \frac{(-1)^{n}\mu_0^2 m_0^2}{4\hbar\pi^3} \int_0^\infty d\omega\, F(\omega, t)\coth{\frac{\hbar\omega}{2k_BT}} \\& \qquad \qquad \qquad \int_0^\infty dq\, q J_{2n}(qD) e^{-2qz} \mathcal{O}^{2n}(q,\omega),
\label{c_dephasometry}
\end{aligned}
\end{align}
where $\Phi_c^{2n}$ and $\mathcal{O}^{2n}$ are the  Fourier coefficients of collective dephasing $\Phi_c(\beta)$ and material response $\mathcal{O}(q,\theta_q)$. 

Equations~(\ref{Phic_Fourier},~\ref{c_dephasometry}) reveal the scheme of correlated quantum dephasometry. Here, the rotation of two-qubit orientation $\beta$ in $\Phi_c(\beta)$ performs angular Fourier analysis of $\mathcal{O}(q,\theta_q)$ in momentum space. In particular, the $2n$th Fourier coefficient $\Phi_c^{2n}$ isolates the $2n$th order rotational symmetry of material response $\mathcal{O}^{2n}$ in momentum space. These even orders $\mathcal{O}^{2n}$ are sufficient to characterize $\mathcal{O}(q,\theta_q)$ of materials with inversion symmetry. Meanwhile, we extend correlated quantum dephasometry to isolate odd orders $\mathcal{O}^{2n+1}$ for materials without inversion symmetry in~\cite{SI}.

Meanwhile, the interqubit distance $D$ and height $z$ decide the sensitivity of the measured signal $\Phi_c$ to target $\mathcal{O}^{2n}$. Here, the near-field factor $qe^{-2qz}$ sets the dominant momentum magnitude $q \sim 1/z$ in the q-integral in Eq.~(\ref{c_dephasometry}), and $D$ sets the weight of $\mathcal{O}^{2n}$ through Bessel function $J_{2n}$ behaviors around $\sim D/z$. Increasing $D/z$ generally improves access to higher-order $\mathcal{O}^{2n}$ by enhancing corresponding $J_{2n}$, while suppressing lower-order ones through increasingly rapid oscillations of lower-order $J_{2n}$ around dominant momentum $1/z$. Systematically varying $D/z$ enables wavevector-resolved reconstruction of material response $\mathcal{O}(q,\theta_q)$~\cite{SI}.

Finally, as in standard quantum noise spectroscopy~\cite{machado2023quantum,cywinski2008enhance}, the measurement pulse sequences (e.g., dynamical decoupling) set the center frequencies $\omega_{dd}$ of the filter function $F(\omega,t)$, further providing the frequency selectivity of $\mathcal{O}(\omega)$. Combining these elements, we have $\Phi_c^{2n} \sim J_{2n}(D/z)\mathcal{O}^{2n}(1/z,\omega_{dd})$. Together, correlated quantum dephasometry can enable tomography of material response resolved in wavevector $(q, \theta_q)$ and frequency $\omega$ with three control knobs, the two-qubit orientation $\beta$, geometric parameter $D/z$, and measurement sequences. 

For comparison, we show that single-qubit quantum dephasometry $\Phi_s(\alpha)$ has limited angular resolution of materials in momentum space. For a single qubit parallel to the 2D material and oriented at angle $\alpha$ from the x-axis [Fig.~\ref{fig:fig1}(a)], we find the single qubit dephasing function $\Phi_s(\alpha,t)=\sum_{n=0,\pm1} e^{i2n\alpha} \, \Phi^{2n}_s(t)$,
\begin{align}
&\begin{aligned}
\Phi^{0,\pm2}_s&= \frac{\mu_0^2 m_0^2}{4\hbar\pi^3} \int_0^\infty d\omega\, F(\omega, t) \coth{\frac{\hbar\omega}{2k_BT}} \\& \quad \quad \frac{1}{3-i^{0,\pm2}} \int_0^\infty dq\, q e^{-2qz} \, \mathcal{O}^{0,\pm2}(q,\omega).
\end{aligned}\label{s_dephasometry}
\end{align}
Comparing Eqs.~(\ref{c_dephasometry},~\ref{s_dephasometry}), single qubit dephasometry accesses only 0th and 2nd order rotational symmetry of $\mathcal{O}(q,\theta_q)$ in momentum space [Fig.~\ref{fig:fig1}(c)], precluding reconstruction of $\mathcal{O}(q,\theta_q)$ and missing material information encoded in higher rotational symmetry channels.

\emph{Isolating Superconductor Pairing Symmetry.--}
As a concrete application, we first apply correlated quantum dephasometry to distinguish s-, d-, and g-wave pairing symmetry of 2D superconductors~\cite{tsuei2000pairing}. Here, the low-frequency ($\sim$MHz) correlated magnetic noise arises from current fluctuations in the superconductor connected to the transverse conductivity $\sigma^T(\mathbf{q},\omega)$ through FDT. In the near-field regime, the correlated noise spectra $J_c$ is dominated by the dissipative response~\cite{dolgirev2022characterizing,sun2025superconducting}, with $\mathcal{O}(\mathbf{q},\omega) = \omega \mathrm{Re} \, \sigma^T(\mathbf{q},\omega)$~\cite{SI}. We note that the rotational symmetry of $\mathrm{Re}\,\sigma^T(\mathbf{q},\omega)$ in momentum space inherits the anisotropy of the superconducting gap $\Delta(\mathbf k)$, enabling direct decomposition of superconductor pairing symmetry with correlated quantum dephasometry. 

We adopt a standard BCS mean-field description for $\mathrm{Re} \, \sigma^T(\mathbf{q},\omega)$ of a 2D superconductor~\cite{dolgirev2022characterizing}. In the Nambu basis, the Bogoliubov-de Gennes (BdG) Hamiltonian is
\begin{equation}\label{BdGH}
    H_{BdG}(\mathbf k)=\varepsilon(\mathbf k) \tau_3 + \Delta(\mathbf k)\tau_1
\end{equation}
where $\tau_i$ are the Pauli matrices in the Nambu space, $\varepsilon_{\mathbf k}$ is the normal state electron dispersion, $\Delta_{\mathbf k}$ is the superconducting gap function. The corresponding retarded BdG Green function is $G^R(\mathbf k, \omega)=\big[ \hbar(\omega+i\Gamma_p)\tau_0 -\varepsilon(\mathbf k)\tau_3 - \Delta(\mathbf k)\tau_1 \big]^{-1}$, where $\Gamma_p$ denotes a phenomenological quasiparticle impurity scattering rate~\cite{dolgirev2022characterizing}. 
The normalized transverse conductivity can be written in terms of the product of BdG Green functions~\cite{dolgirev2022characterizing},
\begin{align}\label{norm_sigma}
\begin{aligned}
&\frac{\mathrm{Re}\,\sigma^T(\mathbf{q},\omega)}{\sigma_n(\bm{0},0)}
=  \frac{(\omega^2+4\Gamma_p^2)}{2\mu\Gamma_p\omega}
\int\frac{d^2 \mathbf{k}}{(2\pi)^2}
\int d\omega_1\;
v_T^2 \, [n_F(\omega_1) \\  & -n_F(\omega_1+\omega)] \mathrm{Tr}\left[ \mathrm{Im}G^R(\mathbf k_-,\omega_1) \mathrm{Im}G^R(\mathbf k_+,\omega_1+\omega)\right],
\end{aligned}
\end{align}
where $\sigma_n$ is the normal state conductivity, $n_F$ is the Fermi-Dirac distribution, $\mathbf{k}_\pm=\mathbf{k}\pm\frac{\mathbf{q}}{2}$, $v_T$ is the transverse electron velocity, and $\mu$ is the chemical potential.

Previous work uses single-qubit relaxometry to probe the $q$-magnitude dependence of $\sigma^T(q)$ to separate nodal from fully gapped pairings in the dirty limit~\cite{dolgirev2022characterizing}. In contrast, we show correlated quantum dephasometry directly resolves the angular $\theta_q$ symmetry of $\sigma^T(q,\theta_q)$, distinguishing different nodal pairings and applicable to the clean regime. From Eq.~(\ref{norm_sigma}), the rotational symmetry of $\mathrm{Re} \, \sigma^T(\mathbf{q},\omega)$ in momentum space is determined by the gap function $\Delta^2(\mathbf{k})$. We consider the model gap functions $\Delta_s=\Delta_0$, $\Delta_d=\Delta_0 \sin{2\theta_k}$, $\Delta_g=\Delta_0\sin{4\theta_k}$, for s-, d-, g-wave superconductors respectively~\cite{tsuei2000pairing}, where $\theta_k$ is the angle of $\mathbf{k}$. Accordingly, the transverse conductivity $\mathrm{Re} \, \sigma^T(\mathbf q,\omega)$ in momentum space is isotropic for s-wave, has four-fold rotational symmetry for d-wave, and eight-fold rotational symmetry for g-wave. In Fig.~\ref{fig:fig2}(a-c), we plot the normalized transverse conductivity $\sigma^T(\mathbf q,\omega)/\sigma_n$ with $\Delta_0/\mu=0.005$, $\hbar\omega/\mu=10^{-7}$, $\hbar\Gamma_p/\mu=5\times10^{-5}$, and $T=0.8T_c$, corresponding to the clean limit. 

In Fig.~\ref{fig:fig2}(d-i), we compare the single-qubit quantum dephasometry and correlated two-qubit dephasometry for resolving pairing symmetry of 2D superconductors. Here, we introduce a characteristic timescale $t_{sc}\approx 2\pi z\hbar \mu/k_F \sigma_n  \Gamma_p m_0^2\mu_0^2 k_BT$ for dephasing functions $\Phi_s(t_{sc}), \Phi_c(t_{sc})$, which roughly corresponds to the single spin dephasing time near s-wave superconductors, and $k_F$ is the Fermi wavevector. As shown in Fig.~\ref{fig:fig2}(d-f), single qubit dephasing $\Phi_s(\alpha)$ exhibits qualitatively similar behaviors near s-, d-, g-wave superconductors. This is because the d-wave and g-wave symmetries are encoded in high order rotational symmetry (4th and 8th) of material response in momentum space difficult to access by single qubit dephasing from Eq.~(\ref{s_dephasometry}). In contrast, correlated dephasing resolves full rotational symmetry of the material response function in momentum space. Therefore, $\Phi_c(\beta)$ exhibits qualitatively different behaviors near s-, d-, g-wave superconductors [Fig.~\ref{fig:fig2}(g-i)], which directly distinguish the superconducting pairing symmetry. 

\begin{figure}[!t]
    \centering
    \includegraphics[width=3.4in]{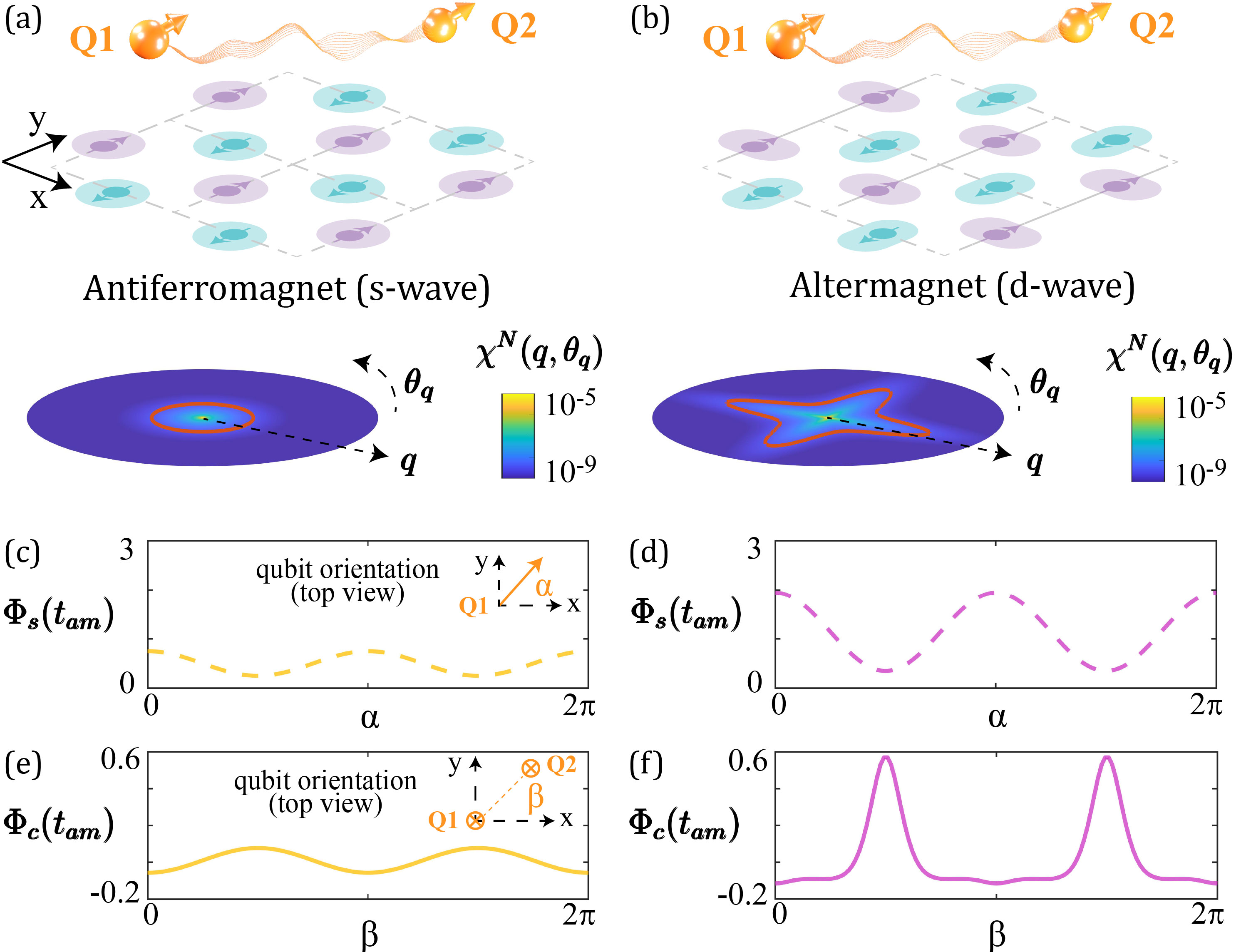}
    \caption{Two-qubit correlated quantum dephasometry resolves s-wave antiferromagnets and d-wave altermagnets. (a,b) Schematic of two spin qubits (Q1, Q2) near 2D antiferromagnets and altermagnets. The rotational symmetry of magnetic susceptibility $\mathrm{Im} \, \chi^N(q,\theta_q)$ in momentum space inherits symmetry of magnon band splittings. (c,d) Single-qubit dephasing function $\Phi_s(\alpha)$ exhibits similar behaviors near s-, d-wave magnets. (e,f) Correlated dephasing function $\Phi_c(\beta)$ 
    exhibits qualitatively distinct angular signatures, directly resolving the d-wave symmetry of 2D altermagnets. We take $D=9z$ in (e,f).}
    \label{fig:fig3}
\end{figure}

\emph{Resolving Altermagnet Symmetry.--}
To highlight the broader applicability, we show the
same framework can distinguish s-wave antiferromagnets and d-wave altermagnets~\cite{vsmejkal2022emerging,vsmejkal2022beyond} for 2D magnets. Here, the correlated magnetic noise probed by spin qubits originates from spin fluctuations connected to the magnetic susceptibility through FDT. At frequencies ($\sim$MHz) deep below the magnon bandgap ($>$GHz)~\cite{wang2022noninvasive}, the dominant contribution to noise correlation spectra $J_c(\omega)$ comes from susceptibility along the Néel axis $\chi^N(\mathbf{q},\omega)$~\cite{wang2022noninvasive}, with $\mathcal{O}(\mathbf{q},\omega) = q^2 \mathrm{Im} \, \chi^N(\mathbf{q},\omega)\cos^2{\theta_q}$~\cite{SI}. Unlike superconductors, Néel axis orientation in real space introduces additional anisotropy $\cos^2{\theta}$, while the symmetry of magnetic order (s-/d-wave) is encoded in the symmetry of $\chi^N(\mathbf{q},\omega)$ in momentum space.  

We follow the Lieb lattice Heisenberg Hamiltonian model for susceptibility of insulating altermagnets and antiferromagnets~\cite{bittencourt2025quantum}. At low frequencies relevant to dephasing noise, $\chi^N$ is governed by spin diffusion dynamics~\cite{wang2022noninvasive,flebus2018quantum,bittencourt2025quantum},
\begin{equation}\label{chi_norm}    \frac{\chi^N(\mathbf{q},\omega)}{\hbar\chi_0\gamma^2}=\frac{\Gamma_m+D(\mathbf{q})}{-i\omega+\Gamma_m+D(\mathbf{q})}
\end{equation}
where $\chi_0$ is the static magnetic susceptibility, $\Gamma_m$ is the magnon relaxation rate, $\gamma$ is the gyromagnetic ratio of spins in the magnet, and $D(\mathbf{q})$ is the effective momentum space spin diffusion coefficient for spin density variations along the Néel axis. 

From Eq.~(\ref{chi_norm}), the rotational symmetry of $\chi^N(q,\theta_q)$ in momentum space is determined by $D(\mathbf{q})$. For s-wave antiferromagnets, $D_s(\mathbf q)=D_0 q^2$. Meanwhile, for d-wave altermagnets, the magnon band splittings~\cite{vsmejkal2022emerging,krempasky2024altermagnetic} generate an anisotropic diffusion coefficient in momentum space $D_d(\mathbf q)=D_0 q^2 - \frac{D_2^2q^4\cos^2{2\theta_{\mathbf q}}}{-i\omega+\Gamma_m+D_0q^2}$~\cite{bittencourt2025quantum}. Consequently, $\mathrm{Im}\chi^N(\mathbf q,\omega)$
is isotropic for antiferromagnets, while has four-fold rotational symmetry for altermagnets in momentum space, as shown in Fig.~\ref{fig:fig3}(a-b), where we take $D_2/D_0=0.9$ and $\omega/\Gamma_m=10^{-3}$.

We now compare the single-qubit and correlated quantum dephasometry for distinguishing 2D antiferromagnets and altermagnets in Fig.~\ref{fig:fig3}(c-f). We consider a characteristic time scale $t_{am}=16\pi\hbar z^2D_0/\mu_0^2k_BT\chi_0\gamma^2m_0^2$ for dephasing functions $\Phi_s(t_{am}), \Phi_c(t_{am})$. In Fig.~\ref{fig:fig3}(c,d), single-qubit dephasing $\Phi_s(\alpha)$ primarily accesses the real space anisotropy imposed by Néel axis, thus exhibiting similar sinusoidal behaviors for s- and d-wave magnets. Previous work leverages the distance $z$ dependence of this sinusoidal amplitude to distinguish antiferromagnets and altermagnets~\cite{bittencourt2025quantum}. In contrast, correlated dephasing $\Phi_c(\beta)$ directly accesses the angular structure of $\chi^N(q,\theta_q)$ encoded in higher-order rotational symmetry, with sinusoidal $\Phi_c(\beta)$ (Fourier expansion $\Phi_c^4=0$) for antiferromagnets and strongly non-sinusoidal $\Phi_c(\beta)$ for altermagnets (Fourier expansion $\Phi_c^4\neq0$), which isolates d-wave symmetry of altermagnets [Fig.~\ref{fig:fig3}(e,f)].

\emph{Experimental considerations.--}
Our proposal can be realized in state-of-the-art experimental platforms, e.g., NV centers in diamond~\cite{rovny2025multi,dolde2013room}. Recent experiments demonstrated intrinsic shallow NVs ($\sim$10 nm depth) with dephasing time exceeding $1\, \mathrm{ms}$ under dynamical decoupling at room temperatures~\cite{lo2025enhancement}, and performed correlated dephasing function readout in controllable qubit pairs separated by $\sim 10–200 \,\mathrm{nm}$ distance with/without entanglement~\cite{rovny2025multi,dolde2013room,le2025wideband}. In addition, intrinsic noise in diamond generally exhibits weak spatial correlations~\cite{kwiatkowski2018decoherence,bradley2019ten}, further enabling direct isolation of correlated dephasing induced by target materials. 

To compare dephasing from target materials against intrinsic diamond noise, we estimate the characteristic dephasing time using realistic 2D/thin-film superconductors and magnets. For superconductors, correlated quantum dephasometry can be applied to highly crystalline thin-film~\cite{ye2012superconducting,tsen2016nature,saito2016highly} or 2D~\cite{cao2018unconventional,oh2021evidence} superconductors in the clean regime. As a representative example, at $T \approx 30\,\mathrm{K}$, we estimate $t_{sc} \sim 850 \,\mu s$ at $10\,\mathrm{nm}$ away from a thin $10\,\mathrm{nm}$ crystalline FeSe film with carrier density $1.8\times 10^{14} \mathrm{cm}^{-2}$ and mobility $39\,\mathrm{cm^2/Vs}$~\cite{sun2014high,wang2015thickness}. For magnets, correlated quantum dephasometry can be applied to magnets with long spin diffusion length. For timescale estimation, we consider an $\alpha$-Fe2O3 thin film of thickness $10\,\mathrm{nm}$ and spin diffusion constant $D_0=8.9\,\mathrm{cm^2/s}$~\cite{wang2022noninvasive}, and find $t_{am} \sim  39 \,\mu s$ at $z=10 \,\mathrm{nm}$ and $T=200\,\mathrm{K}$. Both time scales can be within the reach of ongoing NV capabilities. 

Finally, although we focus on resolving even orders rotational symmetry of material response in momentum space $\mathcal{O}^{2n}$ in the paper, we extend correlated quantum dephasometry to isolate odd orders rotational symmetry $\mathcal{O}^{2n+1}$ for materials without inversion symmetry in Supplemental Material~\cite{SI}. Together, our approach opens a route to symmetry-resolved spectroscopy of material response in momentum space at low frequencies ($\sim$MHz) and nanoscale dimensions for a broad class of quantum materials.

\emph{Acknowledgments.--}
This work was supported by the Defense Advanced Research Projects Agency under the Quantum Materials Engineering using Electromagnetic Fields (QUAMELEON) program.

\nocite{*}
\bibliography{reference}

\end{document}


\preprint{APS/123-QED}

\title{Supplementary Material for `Correlated Quantum Dephasometry: Symmetry-Resolved Noise Spectroscopy of Two-Dimensional Superconductors and Altermagnets'}

\author{Wenbo Sun}
\affiliation{Elmore Family School of Electrical and Computer Engineering, Birck Nanotechnology Center, Purdue University, West Lafayette, Indiana 47907, USA}
\author{Zubin Jacob}
\email{zjacob@purdue.edu}
\affiliation{Elmore Family School of Electrical and Computer Engineering, Birck Nanotechnology Center, Purdue University, West Lafayette, Indiana 47907, USA}

\date{\today}

\maketitle

\section{Near-Field Correlated Dephasing in Multiqubit Sensor Systems}
In this section, we formulate a general framework for near-field correlated dephasing underlying correlated quantum dephasometry. We consider a multiqubit sensor system coupled to spatially correlated fluctuating (noise) magnetic fields $\hat{\mathbf{B}}^{\mathrm{fl}}$ generated by fluctuating currents or magnetizations in a nearby material. Since our focus is pure dephasing, we consider the longitudinal spin-field coupling, $\hat{H}_{\mathrm{int}}=-\sum_i \mathbf{m}_i \hat{\sigma}_i^z \cdot \hat{\mathbf{B}}^{\mathrm{fl}}(\mathbf{r}_i)$, where \(\mathbf{r}_i\), \(\mathbf{m}_i\), and \(\hat{\sigma}_i^z\) denote the position, magnetic moment, and Pauli operator of the \(i\)th qubit. The total Hamiltonian $\hat{H}_{\mathrm{tot}}=\hat{H}_q+\hat{H}_f+\hat{H}_{\mathrm{int}}$ is,
\begin{equation}\label{DHam}
    \hat{H}_{total} = \sum_i \hbar \omega_{i} \hat{\sigma}_i^+ \hat{\sigma}_i^- + \int d^3 \mathbf{r} \int_0^\infty \hbar \omega \hat{\mathbf{f}}^\dagger (\mathbf{r},\omega) \hat{\mathbf{f}}(\mathbf{r},\omega) d\omega- \sum_i  \mathbf{m}_{i}\hat{\sigma}_i^z \cdot \hat{\mathbf{B}}^{\mathrm{fl}}(\mathbf{r_i}),
\end{equation}
where $\hat{\mathbf{f}}^\dagger, \hat{\mathbf{f}}$ are the field creation and annihilation operators, $\omega_i$ are the qubit resonance frequencies. 

As shown in our previous work~\cite{sun2025nanophotonic}, we can derive a time-local quantum master equation from Eq.~(\ref{DHam}) through the time-convolutionless projection operator technique. The dynamics of the multiqubit sensor system are~\cite{sun2025nanophotonic},
\begin{align}\label{ME}
\begin{aligned}
    &\frac{d \rho}{d t} =   -i[\hat{H}_{d},\rho] + \sum_{i} \frac{1}{2} \frac{d \, \Phi_{s}(\mathbf{r_i},t)}{dt}  \Big[\hat{\sigma}_{i}^z \rho \, \hat{\sigma}_{i}^z - \rho \Big] + \sum_{\substack{i \neq j}} \frac{1}{2} \frac{d \, \Phi_{c}(\mathbf{r_i},\mathbf{r_j},t)}{dt} \Big[\hat{\sigma}_{i}^z \rho \, \hat{\sigma}_{j}^z-\frac{1}{2} \hat{\sigma}_{i}^z \hat{\sigma}_{j}^z \rho - \frac{1}{2} \rho \, \hat{\sigma}_{i}^z \hat{\sigma}_{j}^z \Big],
\end{aligned}
\end{align}
where $\rho$ is the density matrix of the multiqubit system, and $\hat{H}_d$ collects unitary terms that do not affect the dephasing structures discussed here~\cite{zou2024spatially}. The second term describes single-qubit dephasing, while the third term describes pairwise correlated dephasing processes. As discussed in the main text, the correlated dephasing dynamics can be extracted from measurements of collective multiqubit coherence in the sensor system~\cite{rovny2022nanoscale,rovny2025multi,le2025wideband}.

From Eq.~(\ref{ME}), the near-field dephasing dynamics of the sensor system are fully determined by the correlated dephasing functions $\Phi_{c}(\mathbf{r_i},\mathbf{r_j},t)$ and single-qubit dephasing functions $\Phi_{s}(\mathbf{r_i},t)$. We have~\cite{sun2025nanophotonic},
\begin{subequations}
\begin{align}    
    &\begin{aligned}
          &\Phi_{c}(\mathbf{r_i},\mathbf{r_j},t) =  \int_0^{\omega_c} d\omega \frac{4\mu_0}{\hbar \pi} \coth{\frac{\hbar \omega}{2 k_B T}} F(\omega, t) J_c(\mathbf{r_i},\mathbf{r_j},\omega)
         \label{td_cdf},
    \end{aligned}\\
    &\begin{aligned}       
          J_c(\omega)  = \frac{\omega^2}{c^2}   \Big[\mathbf{m}_{i} \cdot 
         \frac{\mathrm{Im}  [\overleftrightarrow{G}_m(\mathbf{r_i},\mathbf{r_j},\omega) +\overleftrightarrow{G}^\intercal_m(\mathbf{r_j},\mathbf{r_i},\omega)]}{2} \cdot \mathbf{m}_{j}^\dagger\Big]
         \label{Jc},
    \end{aligned}
\end{align}
\end{subequations} 
and $\Phi_s(\mathbf{r}_i,t)=\Phi_c(\mathbf{r}_i,\mathbf{r}_i,t)$. Here, $\mu_0$ is the vacuum permeability, $T$ is the environment temperature, $\overleftrightarrow{G}_m$ is the near-field dyadic Green function for magnetic fields, and $F(\omega,t)$ is
a frequency filter function controlled by the measurement
pulse sequences. As an example, for Ramsey experiments, $F(\omega,t) = (1-\cos{\omega t})/\omega^2$ centered around $\omega \approx 0$. Meanwhile, for Carr-Purcell-Meiboom-Gill (CPMG) sequences with $n$ $\pi$-pulses, $F(\omega,t)$ is centered around $\omega \approx \pi n/t$~\cite{cywinski2008enhance}. Thus, pulse sequence engineering provides the frequency selectivity of the target material response~\cite{bar2012suppression}. 

Physically, $J_c(\omega)$ represents the spectrum of spatially nonlocal correlations of fluctuating (noise) magnetic fields $\hat{\mathbf{B}}^{\mathrm{fl}}$ at qubit positions $\mathbf{r}_i, \mathbf{r}_j$ near materials. This can be understood from the fluctuation-dissipation theorem for fluctuating magnetic fields, 
\begin{equation}\label{B_FDT}
    \langle \hat{\mathbf{B}}^{\mathrm{fl}}(\mathbf{r_i},\omega) \hat{\mathbf{B}}^{\mathrm{fl},\dagger}(\mathbf{r_j},,\omega) \rangle = \frac{2\pi\hbar \mu_0 \omega^2}{c^2}\coth\!\Big(\frac{\hbar\omega}{2k_BT}\Big)\,\left[\frac{\overleftrightarrow{G}_m(\mathbf{r}_i,\mathbf{r}_j,\omega)-\overleftrightarrow{G}_m^\dagger(\mathbf{r}_j,\mathbf{r}_i,\omega)}{2i}\right].
\end{equation}

For quantum impurity sensing, spin qubit quantum sensors are typically placed very close to materials (e.g., $z<100\,\mathrm{nm}$) to achieve better sensitivity of material properties. In the pure dephasing regime, the relevant frequencies are $\omega\sim\mathrm{MHz}$. Consequently, we are working in an extreme near-field regime with $1/z \gg \omega/c$. Here, the dominant contribution to magnetic fluctuations near conductors and magnets is from the s-polarized evanescent waves~\cite{sun2025superconducting,sun2025nanophotonic,joulain2003definition,volokitin2007near},
\begin{align}
\begin{aligned}\label{anacpd}
    \overleftrightarrow{G}^r_m (\mathbf{r}_i,\mathbf{r}_j,\omega) \approx \frac{i}{8 \pi^2}\int \frac{d^2 \mathbf{q}}{k_z} e^{i \mathbf{q}(\mathbf{r}_i-\mathbf{r}_j)} e^{i k_z(z_i+z_j)} \biggl( \frac{r_{ss}(\mathbf{q})}{k_0^2q^2}\begin{bmatrix} -q_x^2 k_z^2 & -q_x q_y k_z^2 & -q_x k_z q^2\\-q_x q_y k_z^2 & -q_y^2 k_z^2 & -q_y k_z q^2\\q^2 q_x k_z & q^2 q_y k_z & q^4 \end{bmatrix} \biggr),
\end{aligned}
\end{align}
where $k_z=\sqrt{k_0^2-q^2}$, $k_0=\omega/c$, $z_i$ is the distance between the \textit{i}th qubit and material, and $r_{ss}(\mathbf{q})$ is the Fresnel reflection coefficients for s-polarized waves. In this integral, the factor $e^{ik_z(z_i+z_j)}$ introduces a momentum cutoff $\sim 1/z$, and the integral in Eq.~(\ref{anacpd}) is dominated by contributions from $q\sim 1/z \gg k_0=\omega/c$. Consequently, $k_z \approx iq$. 
Substituting Eq.~(\ref{anacpd}) into Eq.~(\ref{Jc}), we find the noise correlation spectra,
\begin{align}
\begin{aligned}       
          J_c(\mathbf{r}_i,\mathbf{r}_j,\omega) =  & \mathrm{Im} \, \Big[ \frac{\mu_0}{16 \pi^2}\int q \, dq \, d\theta_q \, e^{i \mathbf{q}\cdot \mathbf{D}} e^{-q(z_i+z_j)} \mathcal{O}(\mathbf{q}) \biggl( \mathbf{m}_{i} \cdot \begin{bmatrix} \cos^2{\theta_q} & \cos{\theta_q}\sin{\theta_q} & -i\cos{\theta_q} \\ \cos{\theta_q}\sin{\theta_q} & \sin^2{\theta_q} & -i\sin{\theta_q} \\ i\cos{\theta_q} & i\sin{\theta_q} & 1 \end{bmatrix} \cdot \mathbf{m}_{j}^\dagger  \biggr)  \\ & + \frac{\mu_0}{16 \pi^2}\int q \, dq \, d\theta_q \, e^{-i \mathbf{q}\cdot \mathbf{D}} e^{-q(z_i+z_j)} \mathcal{O}(\mathbf{q}) \biggl( \mathbf{m}_{i} \cdot \begin{bmatrix} \cos^2{\theta_q} & \cos{\theta_q}\sin{\theta_q} & i\cos{\theta_q} \\ \cos{\theta_q}\sin{\theta_q} & \sin^2{\theta_q} & i\sin{\theta_q} \\ -i\cos{\theta_q} & -i\sin{\theta_q} & 1 \end{bmatrix} \cdot \mathbf{m}_{j}^\dagger \biggr) \Big]/2
         \label{Jc_O},
\end{aligned}
\end{align} 
where $\mathcal{O}(\mathbf{q})=\mathcal{O}(q,\theta_q)=2q \, r_{ss}(\mathbf{q})/\mu_0$ and $\mathbf{D}=\mathbf{r}_1-\mathbf{r}_j$. As discussed in the main text (also see derivations below), for 2D materials, $\mathcal{O}(\mathbf{q})$ is directly related to the material response in momentum space, e.g., for 2D superconductors, $\mathcal{O}(\mathbf{q})= \omega \mathrm{Re} \, \sigma^T(\mathbf{q},\omega)$, while for 2D altermagnets, $\mathcal{O}(\mathbf{q})=q^2 \,  \mathrm{Im} \, \chi^N(\mathbf{q},\omega)\cos^2{\theta_q}$. 

We note that the above derivation provides a general framework for near-field correlated dephasing. We note that we obtain the same results by using the fluctuation dissipation theorem (FDT) approach, and connect material current or magnetization fluctuations to nearby magnetic-field fluctuations through the Biot-Savart law~\cite{dolgirev2024local} or magnetic dipolar fields~\cite{machado2023quantum}, respectively.

\section{Two-qubit Correlated Dephasometry}\label{SIS2}
In this section, we derive the main text Eqs.~(3-5) for two-qubit correlated dephasometry, which is valid for materials with/without inversion symmetry. We first consider the perpendicular qubit configurations used in the main text, where correlated quantum dephasometry decomposes the angular structure of material response in momentum space in its simplest form. We then prove that this momentum-space symmetry decomposition is not restricted to that particular choice of qubit orientation, but remains valid for arbitrary qubit orientations. In this section, we focus on extracting even orders rotational symmetry of material response $\mathcal{O}^{2n}$ relevant to examples considered in the main text. We extend correlated quantum dephasometry to extract $\mathcal{O}^{2n+1}$, e.g., present in materials without inversion symmetry in Section~\ref{SIS8}.

\subsection{Correlated Quantum Dephasometry with Perpendicular Qubit Orientations}
We begin with two spin qubits perpendicular to the material interface $\mathbf{m}_i=\mathbf{m}_j=(0,0,m_0)$ considered in the main text. For a spin-1/2 particle, we take $m_0=\hbar\gamma_e/2$. In this case, Eq.~(\ref{Jc_O}) reduces to,
\begin{align}
    \begin{aligned}
        J_c(\omega)    =  \frac{\mu_0 m_0^2}{16 \pi^2}\int dqd\theta_q \, q\, e^{-2qz}  \, \cos{( \mathbf{q} \cdot \mathbf{D})} \, \mathrm{Im} \Big[ \mathcal{O}(\mathbf{q})\Big],
    \end{aligned}
\end{align}
where $\mathbf{D}=\mathbf{r}_i-\mathbf{r}_j$. We consider the Jacobi–Anger expansion, 
\begin{equation}
    \mathrm{Re} \, e^{i \mathbf{q}\cdot \mathbf{D}} = \mathrm{Re} \, e^{iqD \cos{(\beta-\theta_q)}} = \sum_{n=-\infty}^\infty i^{2n} J_{2n}(qD)e^{i2n(\beta-\theta_q)}
\end{equation}
where $D=|\mathbf{D}|$, $q=|\mathbf{q}|$, $\beta$ is the angle of $\mathbf{D}$, and $\theta_q$ is the angle of $\mathbf{q}$. $J_{2n}(qD)$ is the \textit{2n}th Bessel function of the first kind. We have,
\begin{align}
    \begin{aligned}
        J_c(\omega)  & 
         = \frac{\mu_0m_0^2}{16 \pi^2}\int dqd\theta_q \, q\, e^{-2qz}  \, \sum_{n=-\infty}^\infty i^{2n} J_{2n}(qD)e^{i2n(\beta-\theta_q)} \, \mathrm{Im} \Big[\mathcal{O}(\mathbf{q})  \Big] \\ & 
         =  \frac{\mu_0m_0^2}{16 \pi^2} \sum_{n=-\infty}^\infty (-1)^{n} e^{i2n\beta} \int dq \, q\, e^{-2qz}  \,  J_{2n}(qD) \int  d\theta_q \, e^{-i2n\theta_q} \, \mathrm{Im} \Big[\mathcal{O}(q,\theta_q)\Big]
    \end{aligned}\label{Jc_perp}
\end{align}
This is Eq.~(3) in the main text. The angular integral over $\theta_q$ decomposes
the material response $\mathcal{O}(q,\theta_q)$ into Fourier coefficients
$\mathcal{O}^{2n}$, while $J_{2n}(qD)$ sets the weight of each order of rotational symmetry in the correlated noise spectrum $J_c$. Substituting this into Eq.~(\ref{td_cdf}), we obtain Eqs.~(4,5) in the main text.

\subsection{Correlated Quantum Dephasometry with Arbitrary Qubit Orientations}
We now prove that the above momentum-space symmetry decomposition is valid for arbitrary qubit orientations. We consider $\mathbf{m}_i= m_0 (\sin{\varphi_i}\cos{\alpha_i},\,\sin{\varphi_i}\sin{\alpha_i},\,\cos{\varphi_i})$, $\mathbf{m}_j= m_0 (\sin{\varphi_j}\cos{\alpha_j},\,\sin{\varphi_j}\sin{\alpha_j},\,\cos{\varphi_j})$. We note that two-qubit orientation $\beta$ rotation (e.g., rotating the spin substrate/target materials) will also change each qubit orientation $\alpha$. Therefore, it is convenient to take $\alpha_i, \alpha_j$ as summation of fixed angles relative to the two-qubit axis plus the two-qubit angle $\beta$, thus $\mathbf{m}_{i,j}= m_0 (\sin{\varphi_{i,j}}\cos{(\alpha_{i,j}+\beta)},\,\sin{\varphi_{i,j}}\sin{(\alpha_{i,j}+\beta)},\,\cos{\varphi_{i,j}})$. With this convention, we find, 
\begin{align}
    \begin{aligned}
        J_c(\omega) &        
        \approx \frac{\mu_0}{16 \pi^2}\int dqd\theta_q \, q \, e^{-2qz} \, \mathrm{Im} \Big[ \mathcal{O}(\mathbf{q}) \Big] \,  \mathbf{m}_{i} \cdot \Big( \cos{(\mathbf{q}\cdot\mathbf{D})} \begin{bmatrix} \cos^2{\theta_q} & \cos{\theta_q}\sin{\theta_q} & 0 \\ \cos{\theta_q}\sin{\theta_q} & \sin^2{\theta_q} & 0 \\ 0 & 0 & 1 \end{bmatrix} \\& \qquad \qquad \qquad \qquad \qquad \qquad \qquad \qquad \qquad \qquad \qquad + \sin{(\mathbf{q}\cdot\mathbf{D})} \begin{bmatrix} 0 & 0 & \cos{\theta_q} \\ 0 & 0 & \sin{\theta_q} \\ -\cos{\theta_q} & -\sin{\theta_q} & 0 \end{bmatrix}\Big) \cdot \mathbf{m}_{j}  \\ &    =  \frac{\mu_0m_0^2}{16 \pi^2}\int dqd\theta_q e^{-2qz} q \, \mathrm{Im} \Big[\mathcal{O}(\mathbf{q})\Big] \,\mathcal{K}(\varphi_i,\alpha_i,\varphi_j,\alpha_j,\theta_q,\beta,q,D),
    \end{aligned}\label{Jc_arb}
\end{align}
which is valid for materials with/without inversion symmetry, and
\begin{align}
\begin{aligned}
\mathcal{K} &= \cos{\big(qD\cos{(\beta-\theta_q)}\big)}\Big[\cos\varphi_i \cos\varphi_j+\cos(\alpha_i+\beta-\theta_q)\cos(\alpha_j+\beta-\theta_q)\sin\varphi_i\sin\varphi_j\Big]\\ & \qquad \qquad \qquad \qquad +\sin{\big(qD\cos{(\beta-\theta_q)}\big)}\Big[\cos\varphi_j\cos(\alpha_i+\beta-\theta_q)\sin\varphi_i-\cos\varphi_i\cos(\alpha_j+\beta-\theta_q)\sin\varphi_j\Big]\\    & =
\cos{\big(qD\cos{(\beta-\theta_q)}\big)} \left(K_0+K_1 \, e^{i(2\beta-2\theta_q)}+K_2 \, e^{-i(2\beta-2\theta_q)}\right) +\sin{\big(qD\cos{(\beta-\theta_q)}\big)}\left(K_3 \, e^{i(\beta-\theta_q)}+K_4 \, e^{-i(\beta-\theta_q)}\right),
\end{aligned}\label{kappa_full}
\end{align}
and 
\begin{subequations}\label{K_coef}
\begin{equation}
K_0 = \cos\varphi_i\cos\varphi_j
+\frac{1}{2}\sin\varphi_i\sin\varphi_j\cos(\alpha_i-\alpha_j)
\end{equation} 
\begin{equation}
K_1 = \frac{1}{4}\sin\varphi_i\sin\varphi_j\,e^{i(\alpha_i+\alpha_j)}, \qquad \quad
K_2 = \frac{1}{4}\sin\varphi_i\sin\varphi_j\,e^{-i(\alpha_i+\alpha_j)},
\end{equation}
\begin{equation}
K_3 = \frac{1}{2}\left(
\sin\varphi_i\cos\varphi_j\,e^{i\alpha_i}
-\cos\varphi_i\sin\varphi_j\,e^{i\alpha_j}
\right), \qquad \quad 
K_4 = \frac{1}{2}\left(
\sin\varphi_i\cos\varphi_j\,e^{-i\alpha_i}
-\cos\varphi_i\sin\varphi_j\,e^{-i\alpha_j}
\right).
\end{equation} 
\end{subequations}
We now apply the Jacobi–Anger expansion,
\begin{equation}
    \mathrm{Re} \, e^{iqD\cos{(\beta - \theta_q)}} = \sum_{n=-\infty}^\infty (-1)^{n} J_{2n}(qD)e^{i2n(\beta - \theta_q)}, \quad 
    \mathrm{Im} \, e^{iqD\cos{(\beta - \theta_q)}} = \sum_{n=-\infty}^\infty (-1)^{n} J_{2n+1}(qD)e^{i(2n+1)(\beta - \theta_q)},
\end{equation}
and we obtain after reindexing,
\begin{subequations}\label{kappa_n}
\begin{equation}
\mathcal{K}=\sum_{n=-\infty}^{\infty}(-1)^n\mathcal{K}_{2n} e^{i2n(\beta-\theta_q)},
\end{equation}
\begin{equation}
\mathcal{K}_{2n}(qD)
=
K_0J_{2n}(qD)
-K_1J_{2n-2}(qD)
-K_2J_{2n+2}(qD)
-K_3J_{2n-1}(qD)
+K_4J_{2n+1}(qD).
\end{equation}
\end{subequations}
Substituting Eq.~(\ref{kappa_n}) into Eq.~(\ref{Jc_arb}), we have,
\begin{align}
    \begin{aligned}
        J_c(\omega) &
           = \mathrm{Im} \Big[ \frac{\mu_0m_0^2}{16 \pi^2}\int dqd\theta_q e^{-2qz} q \, \mathcal{O}(\mathbf{q}) \,\sum_{n=-\infty}^{\infty}(-1)^n\mathcal{K}_{2n}(qD) e^{i2n(\beta-\theta_q)})\Big] \\ &
           = \mathrm{Im} \Big[ \frac{\mu_0m_0^2}{16 \pi^2} \sum_{n=-\infty}^{\infty} (-1)^n e^{i2n\beta} \int dq e^{-2qz} q  \,\mathcal{K}_{2n}(qD) \int d\theta_q  e^{-i2n\theta_q}\mathcal{O}(q,\theta_q)\Big].
    \end{aligned}\label{Jc_arb_resolved}
\end{align}
Comparing with main text Eq.~(3) (Eq.~(\ref{Jc_perp})), the only change is that the simple Bessel weight function $J_{2n}(qD)$ is replaced by a generalized weight function $\mathcal{K}_{2n}(qD)$ determined by the orientations of the two spin qubit sensors. We note that, for perpendicular qubit orientations $\varphi_i=\varphi_j=0$, Eq.~(\ref{kappa_n}) is reduced to $J_{2n}(qD)$. For parallel qubit orientations with $\varphi_i=\varphi_j, \, \alpha_i=\alpha_j$, we have $K_3 = K_4=0$.

Substituting Eq.~(\ref{Jc_arb_resolved}) into Eq.~(\ref{td_cdf}), we further find the correlated dephasing function,
\begin{subequations}\label{Phi_c_arb}
\begin{equation}
\Phi_c(\beta,D,t)=\sum_{n=-\infty}^\infty e^{i2n\beta} \, \Phi^{2n}_c(D,t),
\end{equation}
\begin{equation} 
\Phi^{2n}_c(D)= \frac{(-1)^{n}\mu^2_0 m_0^2}{4\hbar\pi^3} \int_0^\infty d\omega\, F(\omega, t)\coth{\frac{\hbar\omega}{2k_BT}} \int_0^\infty dq\, q \,  e^{-2qz} \, \mathcal{K}_{2n}(qD) \int d\theta_q  e^{-i2n\theta_q}\,\mathrm{Im} \Big[\mathcal{O}(q,\theta_q)\Big].
\end{equation}
\end{subequations}
Equations.~(\ref{Phi_c_arb}) have the same structure as Eqs.~(4,5) in the main text, the \textit{2n}th Fourier component $\Phi_c^{2n}$ still isolates the \textit{2n}th rotational symmetry channel of the material response $\mathcal{O}^{2n}$ in momentum space. The only difference is that, for arbitrary qubit orientations, the Bessel weight function $J_{2n}(qD)$ in the main text is replaced by a more complex weight function $\mathcal{K}_n(qD)$, which is a linear combination of Bessel functions controlled by the directions of the quantization axes of two qubits. Thus, arbitrary qubit orientations do not change which symmetry channels of material response in momentum space are accessible, they only modify how strongly each channel contributes to the measured correlated dephasing signal.

\section{Comparison with Single-Qubit Dephasometry}
In this section, we derive the main text Eqs.~(6) for comparing single-qubit dephasometry with correlated two-qubit dephasometry. In the main text, we consider a single qubit parallel to the material interface for direct comparison with correlated quantum dephasometry. We prove that the limitation of single-qubit dephasometry in accessing rotational symmetry of material response in momentum space is valid for arbitrary
single qubit orientations.

In the two-qubit case Section S2, the global rotation is carried by two-qubit axis direction $\beta$, and the azimuthal angle of single qubit orientation $\alpha_i, \alpha_j$ is fixed relative to that two-qubit axis. In contrast, here, for the single-qubit case, we consider that the global rotation is carried by the azimuthal angle of single qubit orientation $\alpha_i$ itself. In this convention, we consider a single qubit with $\mathbf{m}_i=m_0 (\sin{\varphi_i}\cos{\alpha_i},\,\sin{\varphi_i}\sin{\alpha_i},\,\cos{\varphi_i})$. Taking $D=0$ in Eq.~(\ref{Jc}), we find,
\begin{align}\label{single_dep}
    \begin{aligned}
        J_s(\omega)  &
        = \frac{\mu_0 m_0^2}{16 \pi^2}\int dqd\theta_q e^{-2qz} q \, \mathrm{Im} \Big[ \mathcal{O}(q,\theta_q) \Big] \,\left(K_0'+K_1' \, e^{i(2\alpha_i-2\theta_q)}+K_2' \, e^{-i(2\alpha_i-2\theta_q)}\right)\\ &
        =\frac{\mu_0 m_0^2}{16 \pi^2} \sum_{n=0,\pm 1} e^{i2n\alpha_i} \int dq e^{-2qz} q  \,\mathcal{K}'_{2n}(0) \int d\theta_q  e^{-i2n\theta_q}\mathrm{Im} \, \Big[ \mathcal{O}(q,\theta_q)\Big],
    \end{aligned}
\end{align}
where $\mathcal{K}_0'(0)=\cos^2{\varphi_i}+\frac{1}{2}\sin^2{\varphi_i}$, $\mathcal{K}_2'(0)=\mathcal{K}_{-2}'(0)=\frac{1}{4}\sin^2{\varphi_i}$. The situation where $\varphi_i=\frac{\pi}{2}$ considered in the main text gives the best capability for single-qubit dephasometry to resolve rotational symmetry of material response in momentum space. We further obtain Eqs.~(6,7) in the main text,
\begin{subequations}\label{Phi_s_arb}
\begin{equation}
\Phi_s(\alpha,t)=\sum_{n=0,\pm1} e^{i2n\alpha} \, \Phi^{2n}_s(t),
\end{equation}
\begin{equation} 
\Phi^{0,\pm2}_s= \frac{\mu^2_0 m_0^2}{4\hbar\pi^3} \int_0^\infty d\omega\, F(\omega, t)\coth{\frac{\hbar\omega}{2k_BT}} \int_0^\infty dq\, q \,  e^{-2qz} \, \mathcal{K}'_{0,\pm2}(0) \int d\theta_q  e^{-i2n\theta_q}\,\mathrm{Im} \Big[\mathcal{O}(q,\theta_q)\Big].
\end{equation}
\end{subequations}

\section{Tomography of material response functions in momentum space}
In this section, we show that correlated quantum dephasometry provides a route for using $\Phi_c(\beta)$ to perform momentum space tomography of the material response functions $\mathcal{O}(q,\theta_q,\omega)$. 

We note that the Eq.~(\ref{Phi_c_arb}) naturally separates the tomography of $\mathcal{O}(q,\theta_q,\omega)$ into three components, the $\theta_q$ dependence, $q$ dependence, and frequency $\omega$ dependence. We first expand the response and correlated dephasing function into their Fourier components $\Phi_c^{2n}$ and $\mathcal{O}^{2n}$. We find,
\begin{equation} 
\Phi^{2n}_c(D) \propto (-1)^{n} \int_0^\infty d\omega\, F(\omega, t)\coth{\frac{\hbar\omega}{2k_BT}} \int_0^\infty dq\, q \,  e^{-2qz} \, \mathcal{K}_{2n}(qD) \mathrm{Im} \Big[\mathcal{O}^{2n}(q,\omega)\Big].
\end{equation}
Therefore, the $\beta$ dependence of $\Phi_c$ directly resolves the $\theta_q$ dependence of the material response. For materials with inversion symmetry, even orders $\Phi_c^{2n}$  fully characterize material response in momentum space. Meanwhile, for materials without inversion symmetry, we extend correlated quantum dephasometry to capture $\mathcal{O}^{2n+1}$ in Section~\ref{SIS8}.

Meanwhile, as discussed in Section S1, the frequency dependence is selected by the filter function $F(\omega,t)$, in the same way as in standard noise spectral decomposition under dynamical-decoupling pulse sequences~\cite{machado2023quantum}. Here, dynamical decoupling pulse sequences with a narrow band filter function centered around $\omega_{dd}$ create,
\begin{equation} 
\Phi^{2n}_c(D,\omega_{dd}) \propto (-1)^{n} \int_0^\infty dq\, q \,  e^{-2qz} \, \mathcal{K}_{2n}(qD) \mathrm{Im} \Big[\mathcal{O}^{2n}(q,\omega_{dd})\Big].
\end{equation}

Finally, the $q$ dependence is resolved from the geometric configurations of the two qubit sensors $D/z$. Discretizing $q\to q_m, m=0,1,2,\cdots$ and measuring at a set of geometries $(D_l,z_l), l=0,1,2,\cdots$, we obtain, for each symmetry channel $2n$,
\begin{equation}
\Phi_{c,l}^{2n}
=
\sum_m M_{lm}^{(2n)} O_m^{2n},
\qquad
M_{lm}^{(2n)}
=
A \, q_m e^{-2q_m z_l} K_{2n}(q_m D_l)\,\Delta q_m,
\label{S19}
\end{equation}
where $\Phi_{c,l}^{2n} = \Phi_c^{2n}(D_l,z_l,\omega_{\rm dd})$,  $\mathcal{O}_m^{2n} = O^{2n}(q_m,\omega_{dd})$, and $A$ is a constant. Thus, the radial profile $\mathcal{O}^{2n}(q,\omega_{dd})$ can be reconstructed by inverting the linear system in Eq.~\eqref{S19}. We note that this method can only access $q$ dependence for $q \leq 1/z$, and cannot access higher momentum behaviors of $\mathcal{O}^{2n}(q)$ at $q\gg1/z$. In addition, the radial tomography is controlled by $D/z$, and it is not necessary to perform independent scans of $D$ and $z$.

Combining the angular decomposition of $\Phi_c(\beta)$, the $D/z$ dependent reconstruction of $O^{2n}(q)$, and the frequency selectivity $\omega_{dd}$ from the pulse sequence, correlated quantum dephasometry can resolve the momentum space response functions $\mathcal{O}(q,\theta_q,\omega)$ of quantum materials with inversion symmetry. We extend correlated quantum dephasometry to materials without inversion symmetry in Section~\ref{SIS8}.

\section{Correlated Quantum Dephasometry for 2D superconductors}
In this section, we provide details for the application of correlated quantum dephasometry to resolve superconducting pairing symmetry.

For a 2D conductor with transverse conductivity $\sigma^T(\mathbf{q},\omega)$, the Fresnel reflection coefficient $r_{ss}(\mathbf{q},\omega)$ is~\cite{van2021optical},
\begin{equation}
    r_{ss}(\mathbf{q},\omega)=-\Big(1+\frac{2k_z}{\mu_0\omega\sigma^T(\mathbf{q},\omega)}\Big)^{-1}.
\end{equation}
As discussed above, in the extreme near-field regime relevant to correlated quantum dephasometry, we have $q \gg k_0$ and $k_z \approx iq$. We thus find $r_{ss}(\mathbf{q}) \approx i \mu_0 \omega \sigma^T(\mathbf{q},\omega) /2q$. Therefore,
\[
\mathrm{Im} \, \mathcal{O}(\mathbf{q}) \approx \mathrm{Re} \, \omega \sigma^T(\mathbf{q},\omega). \] 
Substituting into Eq.~(\ref{Phi_c_arb}), we have,
\begin{subequations}\label{Phic_Fourier_sc}
\begin{equation}
\Phi_c(\beta,D,t)=\sum_{n=-\infty}^\infty e^{i2n\beta} \, \Phi^{2n}_c(D,t),
\end{equation}
\begin{equation}
\Phi^{2n}_c(D)= \frac{(-1)^{n}\mu_0^2 m_0^2}{4\hbar\pi^3} \int_0^\infty d\omega\, F(\omega, t)\coth{\frac{\hbar\omega}{2k_BT}} \int_0^\infty dq\, q \,  e^{-2qz} \, \mathcal{K}_{2n}(qD) \int d\theta_q  \, \omega \, \mathrm{Re}[\sigma^T(q,\theta_q,\omega)] \, e^{-i2n\theta_q}.
\end{equation}
\end{subequations}
For two qubits perpendicular to materials $\varphi_i=\varphi_j=0$, we have $\mathcal{K}_{2n}(qD) = J_{2n}(qD)$.

\subsection{Transverse Conductivity of 2D Superconductors}
For the superconductor application of correlated quantum dephasometry, we adopt the standard mean-field BCS Hamiltonian and Kubo formula for the transverse conductivity $\sigma^T(\mathbf{q},\omega)$ of a 2D superconductor following Ref.~\cite{dolgirev2022characterizing}. For a superconductor with superconducting gap function $\Delta(\mathbf{k})$, the Bogoliubov de-Gennes (BdG) Hamiltonian in the Nambu basis is, 
\begin{equation}\label{BdGH}
    H_{BdG}(\mathbf k)=\varepsilon(\mathbf k) \tau_3 + \Delta(\mathbf k)\tau_1
\end{equation}
where $\tau_i$ are the Pauli matrices in the Nambu space, $\varepsilon_{\mathbf k}=\hbar^2k^2/2m-\mu$ is the normal state electron dispersion, $\Delta({\mathbf k})$ is the superconducting gap function. In this paper, we consider the model gap functions, 
\begin{align}
\begin{aligned}    \Delta_s({\mathbf k})&=\Delta_0, \\
    \Delta_d({\mathbf k})&=\Delta_0 \sin{2\theta_k}, \\
    \Delta_g({\mathbf k})&=\Delta_0\sin{4\theta_k},
\end{aligned}
\end{align}
for s-, d-, g-wave superconductor pairing symmetry, respectively. Here, $\theta_k$ is the angle of $\mathbf{k}$ in momentum space. The corresponding retarded BdG Green function is~\cite{dolgirev2022characterizing}, 
\begin{equation}
    G^R(\mathbf k, \omega)=\big[ \hbar(\omega+i\Gamma_p)\tau_0 -\varepsilon(\mathbf k)\tau_3 - \Delta(\mathbf k)\tau_1 \big]^{-1},
\end{equation}
where $\Gamma_p$ denotes a phenomenological quasiparticle impurity scattering rate~\cite{dolgirev2022characterizing}. From the standard Kubo formula, finite-frequency transverse conductivity $\mathrm{Re}\,\sigma^T(\mathbf q,\omega)$ is obtained from the retarded current-current correlator of the paramagnetic current operator. In mean-field BdG theory, this correlator takes the usual bubble form in Nambu space, which can be written in terms of the trace over two spectral functions $A(\mathbf{k},\omega)=-\mathrm{Im} G^R(\mathbf{k},\omega)/\pi$ in the Nambu space~\cite{dolgirev2022characterizing},
\begin{align}\label{condT}
\mathrm{Re} \, \sigma^T(\mathbf q,\omega)
&=
\frac{e^2\pi}{\omega}
\int\frac{d^2k}{(2\pi)^2}
\int d\omega_1\;
v_T^2(\mathbf k)\,
\Tr\!\big[A(\mathbf k_-,\omega_1)A(\mathbf k_+,\omega+\omega_1)\big]\,
\big[n_F(\omega_1)-n_F(\omega+\omega_1)\big],
\end{align}
where $\mathbf{k}_+=\mathbf{k}+\frac{\mathbf{q}}{2}$, $\mathbf{k}_-=\mathbf{k}-\frac{\mathbf{q}}{2}$, and $v_T$ is the transverse velocity. We note that the above equation reduces to the Drude conductivity $\sigma_n$ in the limit of $T \rightarrow T_c$ (i.e., $\Delta_0 \rightarrow 0$) and $\mathbf{q}\rightarrow \mathbf{0}$. In addition, the rotational symmetry of $\sigma^T(\mathbf{q},\omega)$ in momentum space is determined by the gap function $\Delta(\mathbf{k})$.

For numerical evaluations, we use the dimensionless form of Eq.~(\ref{condT}). We normalize the conductivity by the normal state Drude conductivity $\sigma_n$,
\begin{align}\label{sc_cond_norm}
\frac{\mathrm{Re} \, \sigma^T(\tilde{\mathbf q},\tilde{\omega})}{\sigma_n(\mathbf{0},0)}
&\approx \frac{\pi^2(\tilde{\omega}^2+(2\tilde{\Gamma_p})^2)}{2\tilde{\Gamma_p}}
\int\frac{d^2 \tilde{k}}{(2\pi)^2}
\int d\tilde{\omega}_1\;
\Big(\frac{\partial \tilde{\varepsilon}_\mathbf k}{\partial \tilde{\mathbf k}}\Big)_T^2  \,
\Tr\!\big[\tilde{A}(\tilde{\mathbf k}_-,\tilde{\omega}_1)\tilde{A}(\tilde{\mathbf k}_+,\tilde{\omega}+\tilde{\omega}_1)\big]\,
\frac{n_F(\mu\tilde{\omega}_1)-n_F(\mu\tilde{\omega}+\mu\tilde{\omega}_1)}{\tilde{\omega}},
\end{align}
where we consider $\tilde{\omega} = \omega/(\mu/\hbar)$, $\tilde{\Gamma}_p = \Gamma_p/(\mu/\hbar)$, $\tilde{\mathbf{k}} = \mathbf{k}/k_F$, and 
\begin{align}\label{sc_norm_1}
\tilde{\varepsilon}_\mathbf{k}= \tilde{\mathbf{k}}^2-1, \qquad \tilde{E}_k^2=\tilde{\varepsilon}_\mathbf{k}^2+\Delta_\mathbf{k}^2=(\tilde{\mathbf{k}}^2-1)^2+\Delta_\mathbf{k}^2,
\end{align}
\begin{align}\label{sc_norm_2}
\tilde{A}(\tilde{\mathbf{k}},\tilde{\omega})
=
-\frac{\tilde{\Gamma}_p(\tilde{\omega}^2-\tilde{\Gamma}_p^2-(\tilde{{\mathbf{k}}}^2-1)^2-\tilde{\Delta}_{\mathbf{k}}^2)}{\pi\,\tilde{D}(\tilde{\omega})}\,
\begin{pmatrix}1&0\\0&1\end{pmatrix}
+\frac{2\tilde{\omega}\tilde{\Gamma}_p}{\pi\,\tilde{D}(\tilde{\omega})}\,
\begin{pmatrix}
\tilde{\omega}+\tilde{\mathbf{k}}^2-1 & \tilde{\Delta}_{\mathbf{k}}\\
\tilde{\Delta}_{\mathbf{k}} & \tilde{\omega}-\tilde{\mathbf{k}}^2+1
\end{pmatrix},
\end{align}
\begin{align}\label{sc_norm_3}
\tilde{D}(\tilde{\omega})\equiv (\tilde{\omega}^2-\tilde{\Gamma}_p^2-(\tilde{{\mathbf{k}}}^2-1)^2-\tilde{\Delta}_{\mathbf{k}}^2)^2+(2\tilde{\omega}\tilde{\Gamma}_p)^2, \qquad n_F(\omega)=\frac{1}{e^{\hbar\omega/k_BT}+1}.
\end{align}
Substituting Eq.~(\ref{sc_cond_norm}) into Eq.~(\ref{Phic_Fourier_sc}), we have the correlated dephasing function near 2D superconductors,
\begin{align}
\begin{aligned}
\Phi_c(\beta) &= \sum_{n=-\infty}^\infty e^{i2n\beta} \frac{(-1)^{n}\mu_0^2 m_0^2 \sigma_n}{4\hbar\pi^3} \frac{2k_BT}{\hbar} \int_0^\infty d\omega\, F(\omega, t) \int_0^\infty dq\, q \,  e^{-2qz} \, \mathcal{K}_{2n}(qD) \int d\theta_q\, e^{-i2n\theta_q}  \, \frac{\mathrm{Re}[\sigma^T(q,\theta_q,\omega)]}{\sigma_n}\\ &= \sum_{n=-\infty}^\infty e^{i2n\beta} \frac{(-1)^{n}\mu_0^2 m_0^2 \sigma_n}{4\hbar\pi^3} \frac{2k_BT}{\hbar} \frac{k_F^2\hbar}{\mu}\int_0^\infty d\tilde{\omega}\, F(\tilde{\omega}\mu/\hbar, t) \int_0^\infty d\tilde{q}\, \tilde{q} \,  e^{-2\tilde{q}\tilde{z}} \, \mathcal{K}_{2n}(\tilde{q}\tilde{D}) \int d\theta_q\, e^{-i2n\theta_q}  \, \frac{\mathrm{Re}[\sigma^T(\tilde{q},\theta_q,\omega)]}{\sigma_n}.
\end{aligned}
\end{align}
In the main text, we use Eqs.~(\ref{sc_cond_norm}-\ref{sc_norm_3}) to evaluate $\mathrm{Re} \, \sigma^T(\mathbf q,\omega)/\sigma_n$ and correlated dephasing function $\Phi_c$ near 2D superconductors, and obtain the characteristic time scale $t_{sc}$.

\section{Correlated Quantum Dephasometry for 2D altermagnets}

In this section, we provide details for the application of correlated quantum dephasometry to resolve s-wave antiferromagnets and d-wave altermagnets.

At frequencies deep inside the magnon gap, the magnetic susceptibility relevant to dephasometry is governed by the spin diffusion processes, and the susceptibility tensor is dominated by the component along the Neel axis~\cite{flebus2018quantum,wang2022noninvasive,bittencourt2025quantum}. Without loss of generality, we choose the Neel axis direction as the x-axis. To connect this susceptibility to the correlated dephasing kernel, we now derive the Fresnel reflection coefficients $r_{ss}(\mathbf{q},\omega)$ for a 2D magnet located at $z=0$ with susceptibility $\chi^N(\mathbf{q},\omega)$ along the x axis. 

We consider an incident s-polarized plane wave from $z>0$ with in-plane momentum $\mathbf{q}$, and allow for both s- and p-polarized reflected and transmitted fields. The in-plane component of electric fields is, 
\begin{align}
\begin{aligned}
&\mathbf E^+(\boldsymbol\rho,z) =
E_0\Big[
\hat{\mathbf s}\big(e^{-ik_z z}+r_{ss}e^{+ik_z z}\big)
+
\hat{\mathbf q}\,r_{ps}e^{+ik_z z}
\Big]e^{i\mathbf q\cdot\boldsymbol\rho},
\ z>0,
\\ 
&\qquad \quad \quad \mathbf E^-(\boldsymbol\rho,z)=
E_0\Big[
\hat{\mathbf s}\,t_{ss}
+
\hat{\mathbf q}\,t_{ps}
\Big]e^{-ik_z z}e^{i\mathbf q\cdot\boldsymbol\rho},
\ z<0,
\end{aligned} 
\end{align} 
where $\hat{\mathbf{q}}=(\cos{\theta_q},\sin{\theta_q},0)$, $\hat{\mathbf{s}}=(-\sin{\theta_q},\cos{\theta_q},0)$, $\boldsymbol\rho$ are 2D coordinates, $r_{ss}, r_{ps}$ are Fresnel reflection coefficients, and $t_{ss}, t_{ps}$ are transmission coefficients. At the interface $z=0$, electric fields are,
\begin{equation}\label{Efield}
\mathbf E^+=E_0\big[(1+r_{ss})\hat{\mathbf s}+r_{ps}\hat{\mathbf q}\big],
 \qquad \qquad \qquad 
\mathbf E^-=E_0\big[t_{ss}\hat{\mathbf s}+t_{ps}\hat{\mathbf q}\big].
\end{equation} 
Meanwhile, magnetic fields are,
\begin{equation}
\mathbf H^+
=
E_0\Big[
\frac{k_z}{\omega\mu_0}(1-r_{ss})\hat{\mathbf q}
+
\frac{\omega\varepsilon_0}{k_z} r_{ps}\hat{\mathbf s}
\Big],
\qquad \qquad \qquad 
\mathbf H^-
=
E_0\Big[
\frac{k_z}{\omega\mu_0} t_{ss}\hat{\mathbf q}
-
\frac{\omega\varepsilon_0}{k_z} t_{ps}\hat{\mathbf s}
\Big].
\end{equation} 
At $z=0$, the in-plane components of $\mathbf{E}$ and $\mathbf{H}$ fields satisfy the boundary conditions,
\begin{equation}\label{EMBCs}
\mathbf E^+-\mathbf E^-=
-\,i\omega\mu_0\mathbf{M}_{2D}.
\qquad \qquad \qquad 
\mathbf H^+=\mathbf H^-,
\end{equation} 
where $\mathbf{M}_{2D} = M_x
\hat{\mathbf{x}} = \mu_0 \chi^N(\mathbf{q},\omega) \, H_x(z=0)\hat{\mathbf{x}}$. Solving Eqs.~(\ref{Efield}-\ref{EMBCs}), we obtain the Fresnel reflection coefficients,
\begin{equation}
    r_{ss}
=
-\frac{
i k_z\,\mu_0\,\chi^N(\mathbf q,\omega)\cos^2\theta_q
}{
2
-
i\mu_0\,\chi^N(\mathbf q,\omega)\left(
k_z\cos^2\theta_q+\frac{k_0^2}{k_z}\sin^2\theta_q
\right)
}.
\end{equation}
As discussed above, in the extreme near-field regime relevant to dephasometry, we have $q \gg k_0$ and $k_z \approx iq$. As shown in the main text, at large $q$, $\chi^N(\mathbf{q},\omega) \sim q^{-2}$~\cite{flebus2018quantum}. We thus
find $r_{ss}(\mathbf{q}) \approx q\mu_0\chi^N(\mathbf{q},\omega)\cos^2{\theta_q}/2$. Therefore, 
\begin{equation}
    \mathrm{Im}\, \mathcal{O}(\mathbf{q}) = q^2\,\mathrm{Im}\,\chi^N(\mathbf{q},\omega) \cos^2{\theta_q},
\end{equation}
Substituting into Eq.~(\ref{Phi_c_arb}), we have,
\begin{subequations}\label{Phic_Fourier_am}
\begin{equation}
\Phi_c(\beta,D,t)=\sum_{n=-\infty}^\infty e^{i2n\beta} \, \Phi^{2n}_c(D,t),
\end{equation}
\begin{equation} 
\Phi^{2n}_c(D)= \frac{(-1)^{n}\mu_0^2 m_0^2}{4\hbar\pi^3} \int_0^\infty d\omega\, F(\omega, t)\coth{\frac{\hbar\omega}{2k_BT}} \int_0^\infty dq\, q \,  e^{-2qz} \, \mathcal{K}_{2n}(qD) \int d\theta_q  \, q^2 \, \mathrm{Im}[\chi^N(\mathbf{q},\omega)]\cos^2{\theta_q} \, e^{-i2n\theta_q}.
\end{equation}
\end{subequations}
For $\varphi_i=\varphi_j=0$, we have $\mathcal{K}_{2n}(qD) = J_{2n}(qD)$. We note that we obtain the same results by using the fluctuation-dissipation theorem and connecting near-field magnetic field fluctuations to spin fluctuations in 2D magnets using magnetic dipolar fields.

\subsection{Susceptibility of 2D Altermagnets}
For applying correlated quantum dephasometry to distinguish s-wave antiferromagnets and d-wave altermagnets, we adopt the spin diffusion susceptibility for a 2D altermagnet from Ref.~\cite{bittencourt2025quantum}. At frequencies well below the magnon gap, the relevant response is the diffusive dynamics of the spin density along the Néel axis~\cite{wang2022noninvasive}. In this description, the diffusion tensor is determined by the magnon band structures. Therefore, the anisotropic altermagnetic magnon bands generate an additional anisotropic diffusion component beyond the usual isotropic diffusion in antiferromagnets. Accordingly, the low-frequency response contains an isotropic diffusion part $D_0$, already present in an ordinary antiferromagnet, and an anisotropic part $D_2$, which is specific to the altermagnet. In momentum space, the linearized spin diffusion  hydrodynamic equations for the spin density variation along the Néel axis $\delta S_{\parallel}(\mathbf q,\omega) = S_{\parallel}(\mathbf q,\omega)-\hbar\chi_0 h(\mathbf q,\omega)$ are~\cite{bittencourt2025quantum},
\begin{subequations}
\begin{equation}
\big[-i\omega+\Gamma_m+D_0 q^2\big]\delta S_{\parallel}(\mathbf q,\omega)
-\frac{(D_2 q^2\cos 2\theta_q)^2}{-i\omega+\Gamma_m+D_0 q^2}\,\delta S_{\parallel}(\mathbf q,\omega)
=
i\omega\,\hbar\chi_0 h(\mathbf q,\omega),
\label{spin_diff_para}
\end{equation}
\end{subequations}
where $h(\mathbf q,\omega)=\gamma_e B(\mathbf q,\omega)$ is the external bias, $\gamma_e$ is the gyromagnetic ratio for spins in magnets, $\chi_0$ is the static susceptibility, $\Gamma_m$ is the magnon relaxation rate. The corresponding N\'eel axis direction susceptibility follows from the above equation $\chi^N(\mathbf q,\omega) = \gamma_e S_{\parallel}(\mathbf q,\omega)/B(\mathbf q,\omega)$,
\begin{equation}
\chi^N(\bm q,\omega)
=
\hbar\chi_0\gamma_e^2
\frac{\Gamma_m+D(\bm q)}
{-i\omega+\Gamma_m+D(\bm q)}.
\label{SI_chiN_Deff}
\end{equation}
with the effective momentum dependent diffusion kernel for spin density variations along the Neel axis, 
\begin{equation}
D_{s}(\mathbf q)
=
D_0 q^2, \qquad \qquad D_{d}(\mathbf q)
=
D_0 q^2
-
\frac{D_2^2 q^4\cos^2 2\theta_{q}}{-i\omega + \Gamma_m+D_0 q^2}.
\label{SI_Deff_explicit}
\end{equation}
Therefore, the rotational symmetry of
$\chi^N(\mathbf{q},\omega)$ in momentum space is determined by the effective momentum dependent diffusion kernel $D(\mathbf{k})$. 

Substituting Eq.~(\ref{SI_chiN_Deff}) into Eq.~(\ref{Phic_Fourier_am}), we have the correlated dephasing function near 2D altermagnets and antiferromagnets,
\begin{align}
\begin{aligned}
\Phi_c(\beta) &= \sum_{n=-\infty}^\infty e^{i2n\beta} \frac{(-1)^{n}\mu_0^2 m_0^2 \hbar \chi_0 \gamma_e^2}{4\hbar\pi^3} \int_0^\infty d\omega\, F(\omega, t) \frac{2k_BT}{\hbar \omega} \int_0^\infty dq\, q \,  e^{-2qz} \, \mathcal{K}_{2n}(qD) \int d\theta_q\, e^{-i2n\theta_q}  \, \frac{\mathrm{Im}[q^2\chi^N(q,\theta_q,\omega)\cos^2{\theta_q}]}{\hbar \chi_0 \gamma_e^2}\\&= \sum_{n=-\infty}^\infty e^{i2n\beta} \frac{(-1)^{n}\mu_0^2 m_0^2 \chi_0 \gamma_e^2}{4\pi^3} \frac{1}{\Gamma_m l_s^4} \int_0^\infty d\tilde{\omega}\, F(\tilde{\omega}, \tilde{t}) \frac{2k_BT}{\hbar\tilde{\omega}}  \int_0^\infty d\tilde{q}\, \tilde{q}^3 \,  e^{-2\tilde{q}\tilde{z}} \, \mathcal{K}_{2n}(\tilde{q}\tilde{D}) \int d\theta_q\, e^{-i2n\theta_q}  \, \frac{\mathrm{Im}[\chi^N(\tilde{q},\theta_q,\tilde{\omega})\cos^2{\theta_q}]}{\hbar \chi_0 \gamma_e^2},
\end{aligned}
\end{align}
where $\tilde{\omega}=\omega / \Gamma_m$, $\tilde{q}=ql_s$, $z=z/l_s$, and $D=D/l_s$, where $l_s$ is the spin diffusion length. In the main text, we use the above equations to evaluate $\mathrm{Im} \, \chi^N(\mathbf q,\omega)/\hbar\chi_0\gamma_e^2$ and the correlated dephasing function $\Phi_c$ near 2D altermagnets and antiferromagnets, and obtain the characteristic time scale $t_{am}$.

\section{Experimental Considerations}
In this section, we provide more details for experimental considerations discussed in the main text. We note that correlated quantum dephasometry can be potentially realized in existing shallow spin qubit platforms (e.g., NVs) with/without entanglement readout, and is compatible with realistic material parameters for representative superconducting and magnetic systems.

For the superconductor application, we note that crystalline 2D superconductors (e.g., samples grown by molecular beam epitaxy) in the clean regime are highly preferred~\cite{saito2016highly,wang2015thickness,sun2014high}. This is because the correlated quantum dephasometry probes the rotational symmetry of the momentum dependence of $\mathrm{Re} \,\sigma^T(\mathbf{q},\omega)$, the $q$-dependent anisotropy of the conductivity should remain well resolved rather than being smeared by disorders. In addition, cleaner samples can provide larger normal state conductivity $\sigma_n$, leading to more prominent dephasing induced by target materials, favorable for isolating correlated dephasing induced by materials from intrinsic noise in diamond. As a representative example, in the main text, we consider a $10\,\mathrm{nm}$ crystalline FeSe film with carrier density $1.8\times 10^{14} \mathrm{cm}^{-2}$, mobility $39\,\mathrm{cm^2/Vs}$~\cite{sun2014high,wang2015thickness}. Substituting into Eq.~(\ref{Phic_Fourier_sc}), we estimate a characteristic single-qubit dephasing time $\sim850\,\mathrm{\mu s}$ at $T=30 \, \mathrm{K}$ and qubit-material distance $z=10\,\mathrm{nm}$. Meanwhile, the correlated dephasing signal is controlled mainly by the two qubit distance $D$. In practice, $D \sim 8z - 12 z$, comparable to the main text Fig.~2, gives prominent correlated signals with $\Phi_c/\Phi_s \sim 0.2$. Finally, the angular scan of $\Phi_c(\beta)$ may be implemented by rotating the crystalline sample relative to a fixed qubit pair, so that the material symmetry is mapped directly into the measured correlated dephasing signal as a function of rotation angle $\beta$.

For the altermagnet applications, the most favorable systems are antiferromagnetic or altermagnetic films with sufficiently long spin diffusion length. Since correlated quantum dephasometry probes the momentum-space structure of the susceptibility $\mathrm{Im}\,\chi^N(\mathbf{q,\omega})$ at momenta $q \sim 1/z$, one prefers the regime $ql_s \sim l_s/z \gg 1$, so that the q-dependent response is not strongly smeared by magnon impurity scattering. As a representative example, in the main text, we consider a 10 nm $\alpha$-Fe2O3 film with spin diffusion constant $D_0=8.9\,\mathrm{cm^2/s}$ and spin diffusion length $l_s\sim 3\,\mathrm{\mu m}$~\cite{wang2022noninvasive}, and estimate the static susceptibility $\chi_0$ from Ref.~\cite{wang2022noninvasive}. We estimate a characteristic single qubit dephasing time $\sim 39\,\mathrm{\mu s}$ at sensor height $z=10\,\mathrm{nm}$ and temperature $T=200\,\mathrm{K}$ from Eq.~(\ref{Phic_Fourier_am}). This shorter time scale indicates that the magnetic case is substantially easier to access experimentally than the superconducting one. We note that the correlated signal decreases faster with interqubit separation $D$ compared to the superconductor case, but remains observable in the regime used in Fig.~3. For $D=9z$, we still find $|\Phi_c/\Phi_s| \sim 0.17$ for altermagnets. 

\section{Correlated Quantum Dephasometry for Materials without Inversion symmetry}\label{SIS8}
In this section, we show that correlated quantum dephasometry achieves complete angular tomography of material response $\mathcal{O}(q,\theta_q)$ in momentum space, regardless of inversion symmetry. While the correlated dephasing function $\Phi_c(\beta)$ analyzed in the main text isolates even-order rotational harmonics $\mathcal{O}^{2n}$, we now show a complementary correlated phase rotation function $\Psi_c(\beta)$ derived here, isolates odd-order harmonics $\mathcal{O}^{2n+1}$ present in materials with broken inversion symmetry. Together, $\Phi_c(\beta)$ and $\Psi_c(\beta)$ resolve the full angular structure of $\mathcal{O}(q,\theta_q)$.

In the main text, we focus on extracting even orders rotational symmetry of material response in momentum space $\mathcal{O}^{2n}$, sufficient to characterize materials with inversion symmetry, but miss odd orders $\mathcal{O}^{2n+1}$ present in materials without inversion symmetry. We now reveal that correlated quantum dephasometry can be easily extended to extract $\mathcal{O}^{2n+1}$ present in materials without inversion symmetry. We note that, in this case, we generally have $\overleftrightarrow{G}_m(\mathbf{r}_i,\mathbf{r}_j,\omega) \neq \overleftrightarrow{G}^T_m(\mathbf{r}_j,\mathbf{r}_i,\omega)$ in Eq.~(\ref{Jc}). 

From Eqs.~(\ref{Jc},\ref{Jc_arb}), odd orders rotational symmetry $\mathcal{O}^{2n+1}(q,\theta_q)$ of material response in momentum space will not contribute to the $J_c(\omega)$ and the dissipative dynamics related to $\Phi_s,\Phi_c$ in the master equation Eq.~(\ref{ME}). Instead, from our previous work~\cite{sun2025nanophotonic}, we note that these odd orders $\mathcal{O}^{2n+1}$ lead to a nonzero anti-symmetric part of noise correlations in Eq.~(\ref{B_FDT}), $\mathrm{Im} \big[\langle \hat{\mathbf{B}}^{\mathrm{fl}}(\mathbf{r_i},\omega) \hat{\mathbf{B}}^{\mathrm{fl},\dagger}(\mathbf{r_j},,\omega) \rangle \big] \propto \mathrm{Re}\big[-\overleftrightarrow{G}_m(\mathbf{r}_i,\mathbf{r}_j,\omega)+\overleftrightarrow{G}_m^\intercal(\mathbf{r}_j,\mathbf{r}_i,\omega)\big]/2 \neq 0$,  which contribute to the coherent dynamics in the master equation Eq.~(\ref{ME}). Following our derivations in Ref.~\cite{sun2025nanophotonic}, we explicitly separate the unitary dynamics of sensor systems into two parts,
\begin{align}\label{ME2}
\begin{aligned}
    &\frac{d \rho}{d t} =   -i[\hat{H}'_{d},\rho] + \sum_{ij} \frac{1}{2} \frac{d \, \Phi_{c}(\mathbf{r_i},\mathbf{r_j},t)}{dt} \Big[\hat{\sigma}_{i}^z \rho \, \hat{\sigma}_{j}^z-\frac{1}{2} \hat{\sigma}_{i}^z \hat{\sigma}_{j}^z \rho - \frac{1}{2} \rho \, \hat{\sigma}_{i}^z \hat{\sigma}_{j}^z \Big] \\ & \qquad \qquad \qquad \qquad \qquad \qquad \qquad \qquad \qquad \qquad +  i \sum_{ij} \frac{1}{2} \frac{d \, \Psi_{c}(\mathbf{r_i},\mathbf{r_j},t)}{dt} \Big[\hat{\sigma}_{i}^z \rho \, \hat{\sigma}_{j}^z-\frac{1}{2} \hat{\sigma}_{i}^z \hat{\sigma}_{j}^z \rho - \frac{1}{2} \rho \, \hat{\sigma}_{i}^z \hat{\sigma}_{j}^z \Big],
\end{aligned}
\end{align}
where the second term is the dephasing term discussed in the main text. Meanwhile, for materials without inversion symmetry, we note that the nonzero antisymmetric part of near-field noise correlations introduces an additional third term in Eq.~(\ref{ME}) in the unitary dynamics of the multiqubit sensor systems corresponding to correlated phase rotation $\Psi_c$~\cite{sun2025nanophotonic}. Unlike correlated dephasing function $\Phi_c(\mathbf{r}_i,\mathbf{r}_j,t)=\Phi_c(\mathbf{r}_j,\mathbf{r}_i,t)$, this correlated phase rotation function satisfies $\Psi_c(\mathbf{r}_i,\mathbf{r}_j,t)=-\Psi_c(\mathbf{r}_j,\mathbf{r}_i,t)$ and $\Psi_c(\mathbf{r}_i,\mathbf{r}_i,t)=0$. To exemplify effects of the $\Psi_c$ term in quantum dynamics, we find for a two-qubit density matrix $\rho$~\cite{zou2024spatially},
\begin{align}
\frac{d\rho_{12}}{dt} = i \frac{d \, \Psi_{c}}{dt}\,\rho_{12},
\qquad 
\frac{d\rho_{13}}{dt} = -i\frac{d \, \Psi_{c}}{dt}\,\rho_{13},
\qquad
\frac{d\rho_{34}}{dt} = i \frac{d \, \Psi_{c}}{dt}\,\rho_{34},
\qquad 
\frac{d\rho_{24}}{dt} = -i\frac{d \, \Psi_{c}}{dt}\,\rho_{24}
\end{align}
Different from collective dephasing $\Phi_c$, we note that this $\Psi_c$ leads to a correlated phase rotation, which can be measured in state tomography of the multiqubit sensor systems, and $\Psi_c$ does not affect the populations and Bell-state coherence.

Following our derivations in Ref.~\cite{sun2025nanophotonic}, we can find the correlated phase rotation function $\Psi_c$,
\begin{subequations}
\begin{align}    
    &\begin{aligned}
          &\Psi_{c}(\mathbf{r_i},\mathbf{r_j},t) =  \int_0^{\omega_c} d\omega \frac{4\mu_0}{\hbar \pi} \coth{\frac{\hbar \omega}{2 k_B T}} F(\omega, t) \eta_c(\mathbf{r_i},\mathbf{r_j},\omega)
         \label{td_cdf_imag},
    \end{aligned}\\
    &\begin{aligned}       
          \eta_c(\omega)  = \frac{\omega^2}{c^2}   \Big[\mathbf{m}_{i} \cdot 
         \frac{\mathrm{Re}  [-\overleftrightarrow{G}_m(\mathbf{r_i},\mathbf{r_j},\omega) +\overleftrightarrow{G}^\intercal_m(\mathbf{r_j},\mathbf{r_i},\omega)]}{2} \cdot \mathbf{m}_{j}^\dagger\Big]
         \label{Jc_imag}.
    \end{aligned}
\end{align}
\end{subequations}
Similar to Section~\ref{SIS2}, we find
\begin{align}
    \begin{aligned}
        \eta_c(\omega) &
        \approx  \frac{\mu_0}{16 \pi^2}\int dqd\theta_q e^{-2qz} q \, \mathrm{Im} \Big[  \mathcal{O}(\mathbf{q}) \Big]  \,  \mathbf{m}_{i} \cdot \Big( \sin{(\mathbf{q}\cdot\mathbf{D})} \begin{bmatrix} \cos^2{\theta_q} & \cos{\theta_q}\sin{\theta_q} & 0 \\ \cos{\theta_q}\sin{\theta_q} & \sin^2{\theta_q} & 0 \\ 0 & 0 & 1 \end{bmatrix} \\& \qquad \qquad \qquad \qquad \qquad \qquad \qquad \qquad \qquad \qquad \qquad - \cos{(\mathbf{q}\cdot\mathbf{D})} \begin{bmatrix} 0 & 0 & \cos{\theta_q} \\ 0 & 0 & \sin{\theta_q} \\ -\cos{\theta_q} & -\sin{\theta_q} & 0 \end{bmatrix}\Big) \cdot \mathbf{m}_{j}
        \\& =  \frac{\mu_0m_0^2}{16 \pi^2}\int dqd\theta_q e^{-2qz} q \, \mathrm{Im} \Big[ \mathcal{O}(\mathbf{q}) \Big] \,\mathcal{L}(\varphi_i,\alpha_i,\varphi_j,\alpha_j,\theta_q,\beta,q,D).
    \end{aligned}\label{Jc_no_inv}
\end{align}
where,
\begin{align}
\begin{aligned}
\mathcal{L} &= \sin{\big(qD\cos{(\beta-\theta_q)}\big)}\Big[\cos\varphi_i \cos\varphi_j+\cos(\alpha_i+\beta-\theta_q)\cos(\alpha_j+\beta-\theta_q)\sin\varphi_i\sin\varphi_j\Big]\\ & \qquad \qquad \qquad \qquad -\cos{\big(qD\cos{(\beta-\theta_q)}\big)}\Big[\cos\varphi_j\cos(\alpha_i+\beta-\theta_q)\sin\varphi_i-\cos\varphi_i\cos(\alpha_j+\beta-\theta_q)\sin\varphi_j\Big]\\    & =
\sin{\big(qD\cos{(\beta-\theta_q)}\big)} \left(L_0+L_1 \, e^{i(2\beta-2\theta_q)}+L_2 \, e^{-i(2\beta-2\theta_q)}\right) -\cos{\big(qD\cos{(\beta-\theta_q)}\big)}\left(L_3 \, e^{i(\beta-\theta_q)}+L_4 \, e^{-i(\beta-\theta_q)}\right),
\end{aligned}\label{kappa_full}
\end{align}
and $L_{0,1,2,3,4}=K_{0,1,2,3,4}$ defined in Eq.~(\ref{K_coef}). With the Jacobi-Anger expansion, we thus find after reindexing,
\begin{subequations}\label{L_n}
\begin{equation}
\mathcal{L}=\sum_{n=-\infty}^{\infty}(-1)^n\mathcal{L}_{2n+1} e^{i(2n+1)(\beta-\theta_q)},
\end{equation}
\begin{equation}
\mathcal{L}_{2n+1}(qD)
=
L_0J_{2n+1}(qD)
-L_1J_{2n-1}(qD)
-L_2J_{2n+3}(qD)
-L_3J_{2n}(qD)
+L_4J_{2n+2}(qD),
\end{equation}
\end{subequations}
where $J_n$ are Bessel functions of the first kind. Substituting this into Eq.~(\ref{Jc_arb}), we have,
\begin{align}
    \begin{aligned}
        \eta_c(\omega) &
           = \frac{\mu_0m_0^2}{16 \pi^2} \sum_{n=-\infty}^{\infty} (-1)^n e^{i(2n+1)\beta} \int dq e^{-2qz} q  \,\mathcal{L}_{2n+1}(qD) \int d\theta_q  e^{-i(2n+1)\theta_q} \,  \mathrm{Im} \Big[\mathcal{O}(q,\theta_q)\Big].
    \end{aligned}\label{Jc_arb_resolved_noinv}
\end{align}
We further find the $\Psi_c$,
\begin{subequations}\label{Psi_c_arb}
\begin{equation}
\Psi_c(\beta,D,t)=\sum_{n=-\infty}^\infty e^{i(2n+1)\beta} \, \Psi^{2n+1}_c(D,t),
\end{equation}
\begin{equation} 
\Psi^{2n+1}_c(D)= \frac{(-1)^{n}\mu^2_0 m_0^2}{4\hbar\pi^3} \int_0^\infty d\omega\, F(\omega, t)\coth{\frac{\hbar\omega}{2k_BT}} \int_0^\infty dq\, q \,  e^{-2qz} \, \mathcal{L}_{2n+1}(qD) \int d\theta_q  e^{-i(2n+1)\theta_q}\,\mathrm{Im} \Big[\mathcal{O}(q,\theta_q)\Big].
\end{equation}
\end{subequations}
Therefore, similar to the main text results, the $2n+1$th Fourier coefficient of the correlated phase rotation function $\Psi_c^{2n+1}$ isolates the $2n+1$th order rotational symmetry of material response $\mathcal{O}^{2n+1}$ in momentum space. We note this can be helpful for probing materials with broken inversion symmetry.

Combined with our results in the main text, we emphasize that the angular $\beta$ dependence of correlated dephasing function $\Phi_c(\beta)$ and correlated phase rotation function $\Psi_c(\beta)$ together resolves the full angular structure of material response $\mathcal{O}(q,\theta_q)$ in momentum space, regardless of whether the materials have inversion symmetry or not.

\bibliography{supplemental_reference}